\documentclass[pdflatex,sn-mathphys-num,twocolumns]{sn-jnl}

\usepackage{placeins}
\usepackage{float}
\usepackage{soul}
\usepackage{graphicx}%
\usepackage{multirow}%
\usepackage{amsmath,amssymb,amsfonts}%
\usepackage{amsthm}%
\usepackage{mathrsfs}%
\usepackage[title]{appendix}%
\usepackage{xcolor}%
\usepackage{textcomp}%
\usepackage{manyfoot}%
\usepackage{booktabs}%
\usepackage{algorithm}%
\usepackage{algorithmicx}%
\usepackage{algpseudocode}%
\usepackage{listings}%

\theoremstyle{thmstyleone}%
\theoremstyle{thmstyletwo}%

\theoremstyle{thmstylethree}%

\begin{document}

\title[Article Title]{Machine Learning Accelerated Descriptor Design for Catalyst Discovery in CO$_2$ to Methanol Conversion}


\author[1,2,5]{\fnm{Prajwal} \sur{Pisal}}\email{prajwal.pisal@aalto.fi}
\equalcont{These authors contributed equally to this work.}

\author[1,3]{\fnm{Ond\v{r}ej} \sur{Krej\v{c}\'{i}}}\email{ondrej.krejci@aalto.fi}
\equalcont{These authors contributed equally to this work.}

\author*[1,2,4,5]{\fnm{Patrick} \sur{Rinke}}\email{patrick.rinke@aalto.fi}

\affil[1]{\orgdiv{Department of Applied Physics}, \orgname{Aalto University}, \orgaddress{\street{P.O. Box 11000 }, \city{AALTO}, \postcode{FI-00076}, \country{Finland}}}

\affil[2]{\orgdiv{Department of Physics}, \orgname{Technical University of Munich}, \orgaddress{\street{James-Franck-Strasse 1}, \city{Garching}, \postcode{85748}, \country{Germany}}}

\affil[3]{\orgdiv{Department of Mechanical and Materials Engineering}, \orgname{University of Turku}, \orgaddress{\street{Vesilinnantie 5}, \city{Turku}, \postcode{FI-20014}, \country{Finland}}}

\affil[4]{\orgdiv{Atomistic Modeling Center}, \orgname{Munich Data Science Institute, Technical University of Munich}, \orgaddress{\street{Walther-Von-Dyck Str. 10}, \city{Garching}, \postcode{85748}, \country{Germany}}}

\affil[5]{\orgname{Munich Center for Machine Learning (MCML), \orgaddress{\street{Arcistrasse 21}, \city{Munich}, \postcode{80333}, \country{Germany}}}}



\abstract{

Transforming CO$_2$ into methanol represents a crucial step towards closing the carbon cycle, with thermoreduction technology nearing industrial application. However, obtaining high methanol yields and ensuring the stability of heterocatalysts remain significant challenges. Herein, we present a sophisticated computational framework to accelerate the discovery of thermal heterogeneous catalysts, using machine-learned force fields. We propose a new catalytic descriptor, termed adsorption energy distribution, that aggregates the binding energies for different catalyst facets, binding sites, and adsorbates. The descriptor is versatile and can be adjusted to a specific reaction through careful choice of the key-step reactants and reaction intermediates. By applying unsupervised machine learning and statistical analysis to a dataset comprising nearly 160 metallic alloys, we offer a powerful tool for catalyst discovery. We propose new promising candidates such as ZnRh and ZnPt$_3$, which to our knowledge, have not yet been tested, and discuss their possible advantage in terms of stability.

}




\maketitle

\section{Introduction}\label{sec1}

Utilizing CO$_2$ in the production of useful chemicals closes the carbon loop and subsequently reduces CO$_2$ emissions. 
Converting CO$_2$ into liquid fuels or chemical feedstocks like methanol can decrease our dependence on fossil fuels \cite{ye2019co2}. The hydrogenation of CO$_2$ to methanol involves the reaction of two gases, similar to other important chemical processes like the Haber-Bosch synthesis \cite{ROHR201933}, which produces ammonia, a precursor for fertilizers, from hydrogen and nitrogen. Both processes occur in thermochemical reactors and face significant energetic barriers, requiring high temperatures and pressures to yield the desired products. Heterogeneous catalysis is a key method to lower these reaction barriers, making the processes both technologically and economically viable \cite{96CO2_methanol, ROHR201933}. However, the economic feasibility of methanol synthesis has not yet been achieved \cite{nyari2020techno}.

The identification of an ideal CO$_2$ conversion method focuses primarily on two technological pathways: thermochemical~\cite{ganesh2014renewable} and electrochemical~\cite{Tran2018el-chem}. The thermochemical approach offers significant potential for rapid industrial adoption due to its resemblance to syngas conversion~\cite{ganesh2014renewable}.
Current catalysts, typically based on the industrial syngas catalyst  Cu/ZnO/Al$_2$O$_3$, suffer from low conversion rates, low selectivity~\cite{wang2022co2}, and oxidation poisoning~\cite{li2023deactivation}.
Addressing these issues with better catalysts could increase performance and reduce  costs~\cite{nyari2020techno}.
However, experimentally screening materials to discover effective catalysts remains challenging due to the slow and expensive nature of catalyst testing and the vastness of the materials space.

Computational methods such as density functional theory (DFT) could provide a complementary, efficient and cost-effective alternative for catalyst discovery. However, calculating turn-over frequencies based on reaction barriers is often computationally intensive because it requires explicit transition state calculations.
Moreover, many other factors like catalyst microstructure, reactor designs, mass and heat flows, etc. affect catalytic performance and necessitate multi-scale modeling  \cite{Bruix2019,posada2021firstprinciples}.
Consequently, approximate methods and concepts, such as the Sabatier principle that relate catalytic activity to the adsorption energies of reaction intermediates calculated using DFT~\cite{CHE2013162}, have been frequently employed in extensive searches for candidate materials \cite{medford2015sabatier, Chen2024, CHE2013162}.
Over the years, numerous approximations have been developed that have guided catalyst search, extending the Sabatier principle to correlate activity with more easily obtainable activity descriptors, such as \textit{d}-band center and scaling relations \cite{Jones_2008,vogt2022,medford2015sabatier}.
While these descriptors have provided valuable insight, their usefulness is often constrained to certain surface facets of material or a limited number of material families, such as \textit{d}-metals.

Machine learning (ML) has recently emerged as a powerful alternative in materials research and is gaining traction, with several disciplines employing these techniques to expedite the discovery of new materials~\cite{Ward2016, Ulissi2017, Andersen2019, Himanen/Geurts/Foster/Rinke:2019, zhang2021, Lan2023, Chen2024, mou2023machine, Zhang2018strategy}. 
Data-driven algorithms can analyze vast datasets of catalyst properties and performance, identifying complex relationships that may be beyond the reach of traditional descriptors~\cite{Mamun2020, Fiedler2023}.
In the realm of heterogeneous catalysis, the ML methods can primarily be divided into two categories: mapping catalyst activity using new approximate descriptors \cite{Andersen2019, zhang2021, mou2023machine, Lunger2024} and prediction of adsorption energies using machine-learned force fields (MLFF) \cite{Ulissi2017, Lan2023}.

Both of these approaches have significantly boosted effectiveness in heterogeneous catalyst research~\cite{LIU202425}. Descriptor-based approaches such as SISSO \cite{Andersen2019} and the latest generation of MLFFs based on graph neural networks \cite{Lan2023}, offer remarkable precision. However, these methods require training, which results in a computational cost comparable to direct DFT calculations.

In contrast, pre-trained MLFFs, that make use of extensive DFT datasets, offer explicit relaxation of adsorbates on catalyst surfaces and a significant speed-up (a factor of 10$^4$ or more) compared to DFT calculations while maintaining quantum mechanical accuracy \cite{kang2020, chen2023}, making them suitable for high-throughput catalyst discovery workflows. 
Modern Sabatier principle-based approaches also utilize MLFFs or specialized models to find (global) minimum adsorption energies across multiple material facets \cite{Lan2023, Chen2024} providing a scalable and transferable framework for screening a broad range of catalysts. 
Although several studies emphasize the importance of various catalyst facets \cite{Bruix2019, amann2022}, these approaches predominantly utilize data from individual facets for material characterization. Consequently, the challenge persists: how can we effectively predict catalytic performance without limiting our scope to specific material families or facet orientations?

The absence of an adequate descriptor for the activity of complex materials motivates our exploration of methods to better represent contemporary industrial catalysts. These catalysts, composed of nanostructures with diverse surface facets and adsorption sites, present significant challenges to understanding their performance. This work seeks to address these challenges by focusing on three critical objectives: (1) developing a novel descriptor that captures the structural and energetic complexity of catalysts, (2) establishing an efficient workflow for large-scale computational screening, and (3) devising a robust framework to identify promising candidates from the resulting data.

Firstly, we seek to define a descriptor that encapsulates the inherent complexity of heterocatalytic materials. In this study, we introduce adsorption energy distributions (AEDs) as a tool to represent the spectrum of adsorption energies across various facets and binding sites of nanoparticle catalysts.
Building on recent advances in characterizing structurally complex materials, such as high-entropy alloys \cite{batchelor2019high, pedersen2020high}, we explore the potential of AEDs to fingerprint the material catalytic properties, using the CO$_2$ to methanol conversion reaction as a case study.

Secondly, we aim to establish a high-throughput and ML-enhanced workflow to accelerate the screening of catalytic materials using our newly formulated descriptor. Traditional density functional theory (DFT) approaches are computationally prohibitive for large-scale studies. To overcome this limitation, we leverage MLFFs from the Open Catalyst Project (OCP) \cite{chanussot2021open, tran2023open}, enabling the rapid and accurate computation of adsorption energies. Our workflow generates an extensive dataset of AEDs, capturing over 877,000 adsorption energies for nearly 160 materials relevant to the CO$_2$ to methanol conversion reaction. 
To target the enhanced reliability of our predictions and the effective use of ML models in catalyst discovery, we design a robust validation protocol.

Finally, we address the objective of developing a method to compare AED descriptors, which encode the energy landscape of materials, to those of known effective catalysts. This involves employing unsupervised learning techniques to analyze the extensive dataset of AEDs generated in this study. By treating AEDs as probability distributions, we quantify their similarity using the Wasserstein distance metric \cite{ramdas2017wasserstein} and perform hierarchical clustering to group catalysts with similar AED profiles. This approach enables us to systematically compare the AEDs of new materials to those of established catalysts, identifying potential similarities that suggest comparable performance. Through this comparison with known effective catalysts, we seek to identify new materials with similar AEDs, highlighting a few promising candidates for further investigation.

\section{Results and Discussion}\label{resdis}

\begin{figure}[h]
	\centering 
    \includegraphics{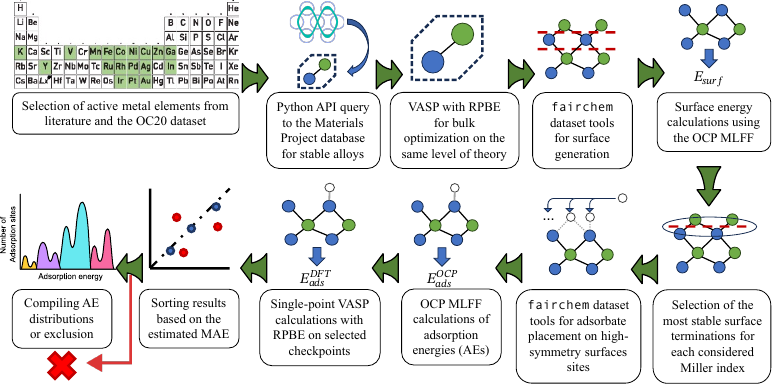}	
	\caption{\textbf{Schematics of the workflow for adsorption energy distributions (AEDs) generation:}
  The AED catalyst database was created through a series of steps, including the choice of metals, bulk optimization, selection of relevant surface geometries, preparation of adsorbate geometries, validation and compilation of AEDs as elaborated in the figure.}
	\label{fig_workflow}
\end{figure}

In order to discover potential new catalysts for converting CO$_2$ into methanol, we present the workflow depicted in Fig.~\ref{fig_workflow}. The key steps are summarized here, with detailed implementation procedures and configurations available in the Methods section~\ref{methods}.

\subsubsection*{Search Space Selection:} 
To effectively reduce the search space for potential catalyst materials for CO$_2$ thermal conversion, we first isolated the metallic elements that have undergone prior experimentation for this process, as documented by Bahri \textit{et al.} ~\cite{bahri2022meta-analysis}. To maintain the prediction accuracy these elements also had to be part of the Open Catalyst 2020 (OC20) database~\cite{chanussot2021open}. The elements shortlisted are the following: K, V, Mn, Fe, Co, Ni, Cu, Zn, Ga, Y, Ru, Rh, Pd, Ag, In, Ir, Pt, and Au. We then proceeded to search through the Materials Project database \cite{matproj2013} for stable and experimentally observed crystal structures associated with these metals and their bimetallic alloys. We compiled 216 stable phase forms involving both single metals and bimetallic alloys corresponding to our set of 18 elements. A detailed listing of these materials is provided in Tables~S1 and S2 of the Supplementary Information. We performed bulk DFT optimization at the RPBE \cite{PhysRevB.59.7413} level to align with the OC20 for the obtained materials. 
Optimization of 22 materials was not successful, and therefore, they were excluded from the materials list, as detailed in Table~S2 in the Supplementary Information~\ref{supple}.

To identify the most crucial adsorbates for AEDs calculations, we perused the existing literature. An experimental investigation by Amman \textit{et al.}~\cite{amann2022} highlighted the presence of surface-bound radicals such as *H (hydrogen atom), *OH (hydroxy group), *OCHO (formate), and *OCH$_3$ (methoxy) as essential reaction intermediates in the thermocatalytic reduction of CO$_2$ to methanol. Based on these findings, we selected these adsorbates for our AEDs calculations. 
Please note that the notation for formate can vary across the literature, e.g., *HCOO~\cite{amann2022,D1CY00922B} and HCOO*~\cite{Wu2017}.
With the help of \texttt{fairchem} repository tools by OCP \cite{fairchem}, we created surfaces with their Miller index $\in \{-2,-1, ... , 2\}$ and calculated their total energy using OCP MLFF.
If we encountered multiple cuts for the same facet, we selected the one with lowest energy for further calculations. 
Then we engineered surface-adsorbate configurations for the most stable surface terminations across all facets within our defined Miller index range for the materials, as described in section \ref{slabgen}, and optimized these configurations using the OCP MLFF.
During this process, we discovered that seven materials exhibited so large surface-adsorbate supercells, that their calculations were infeasible on available GPU resources, even with the effective OCP MLFF.
Consequently, they were excluded from our study.

\subsubsection*{Validation and Data Cleaning:}

\begin{figure}[t!]
	\centering 
	\includegraphics{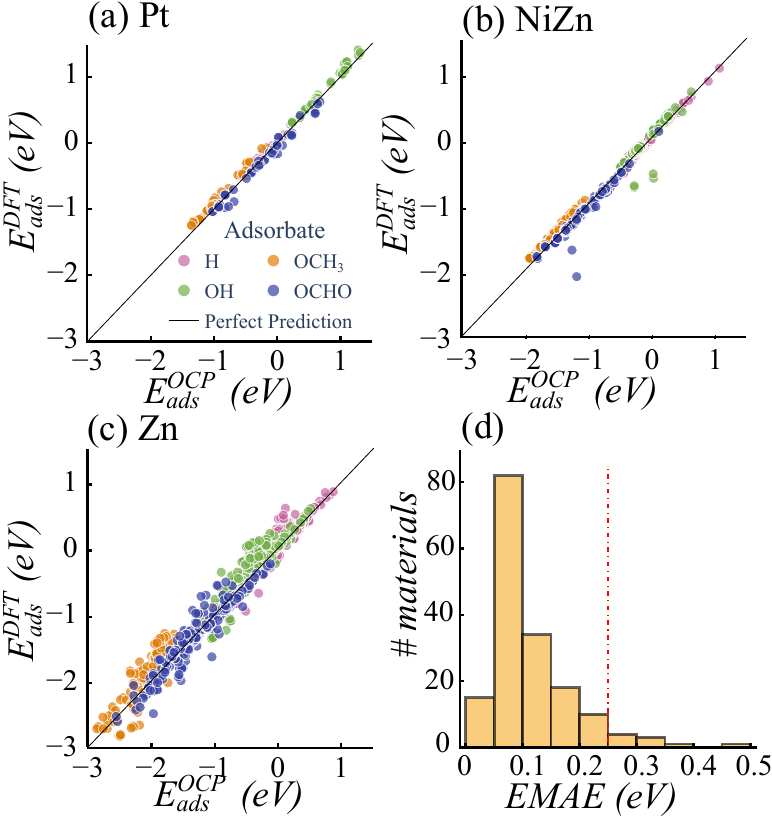}
    \caption{\textbf{Validation of results:}(a-c) The comparison of adsorption energy predicted by the OCP \texttt{equiformer\_V2} MLFF against single-point DFT calculations, for (a) Pt, (b) Zn and (c) NiZn.
    (d) Histogram of EMAEs for the materials that has been calculated with the OCP MLFF, with the 0.25~eV cut-off line, showing that majority of the materials has their EMEA between 0.05 -- 0.10~eV. Out of a total of 188 OCP calculated materials, 17 are not shown here as their EMAE is above 0.5~eV.}
    \label{fig_MAE}
\end{figure}

In our work, we have employed the OCP \texttt{equiformer\symbol{95}V2} MLFF. Its reported accuracy for the adsorption energy of small molecular fragments is 0.23~eV \cite{liao2023equiformerv2}.
However, *OCHO was not included in the OC20 database used for training the \texttt{equiformer\symbol{95}V2}, raising concerns about the accuracy of our adsorption energy predictions in this work. To benchmark \texttt{equiformer\symbol{95}V2} for our use case, we chose Pt, Zn, and NiZn and performed explicit DFT calculations (see Methods section for details).
The comparison between predicted and DFT calculated adsorption energies can be found in Figure~\ref{fig_MAE} and Table~\ref{table_MAE}:
The predictions for Pt are precise, whereas the NiZn results show some outliers, and there is a noticeable degree of scatter for Zn. Despite this, the overall mean absolute error (MAE) for the adsorption energies of the selected materials is 0.16 eV, which is quite impressive and falls within the reported accuracy for the employed MLFF.

To affirm the reliability of our predicted AEDs across a broader range of materials along with maintaining computational practicality, we integrated a validation step within our analysis workflow.
We sampled the minimum, maximum, and median adsorption energies for each adsorbate-material pair from the predicted AEDs.
We performed single-point DFT calculations on these selected systems and compared with the adsorption energy predictions of the OCP MLFF.
The difference is compiled in an `estimated MAE' (EMAE).
Comparisons between EMAE and the all-encompassing MAE for our complete test set are presented in Table~\ref{table_MAE}.
While the EMAE may differ from the actual MAE by up to a factor of three for specific adsorbates, it generally remains in close proximity to the actual MAE, thus serving as a reliable gauge of data quality. 

The validation step is connected with the final data-cleaning when we exclude any material with an EMAE surpassing the threshold of 0.25~eV.
Consequently, 29 materials were expunged from our dataset, retaining 158 materials. Most materials flagged for significant EMAEs exhibited magnetic properties, exemplified by materials like MnCo, MnGa, or FeCo. Magnetism presents significant challenges for the non-spin-polarized DFT calculations used in OC20 and in this work.
A complete list of estimated MAEs for the remaining 158 materials is accessible in \cite{dataset}.

\begin{table}[h]
    \centering
    \caption{Mean absolute error (MAE) obtained through single-point comparison with DFT calculations for three selected materials -- Pt, Zn and NiZn alloy. The MAE is compared to the estimated MAE (EMAE), which is obtained from single-point DFT calculations for only three selected structures per absorbate-material combination.
    }
    \label{table_MAE}
    \begin{tabular}{lccccc}
        \toprule
        Material: & Pt \\
        Adsorbate & *H & *OH & *OCHO & *OCH$_3$ & \textbf{Overall}  \\
        \midrule
        MAE (eV) & 0.02 & 0.05 & 0.06 & 0.09 & \textbf{0.06} \\
        EMAE (eV) & 0.02 & 0.07 & 0.04 & 0.10 & \textbf{0.06}  \\
        MAE/EMAE & 0.97 & 0.66 & 1.59 & 0.93& \textbf{0.96}  \\
        \midrule
        Material: & Zn \\
        Adsorbate & *H & *OH & *OCHO & *OCH$_3$ & \textbf{Overall} \\
        \midrule
        MAE (eV) & 0.10 & 0.13 & 0.10 &  0.15 & \textbf{0.12} \\
        EMAE (eV) & 0.06 & 0.10 & 0.06 & 0.08 & \textbf{0.07} \\
        MAE/EMAE & 1.68 & 1.34 & 1.68 & 1.95 & \textbf{1.64} \\
        \midrule
        Material: & NiZn \\
        Adsorbate & *H & *OH & *OCHO & *OCH$_3$ & \textbf{Overall} \\
        \midrule
        MAE (eV)  & 0.02 & 0.07 & 0.09 & 0.06 & \textbf{0.06} \\
        EMAE (eV) & 0.02 & 0.05 & 0.09 & 0.02 & \textbf{0.05} \\
        MAE/EMAE & 0.73 & 1.36 & 1.04 & 2.61 & \textbf{1.27} \\
        \bottomrule
    \end{tabular}
\end{table}

\subsubsection*{Adsorption Energy Distributions:}
\begin{figure}[t!]
	\centering 
	\includegraphics{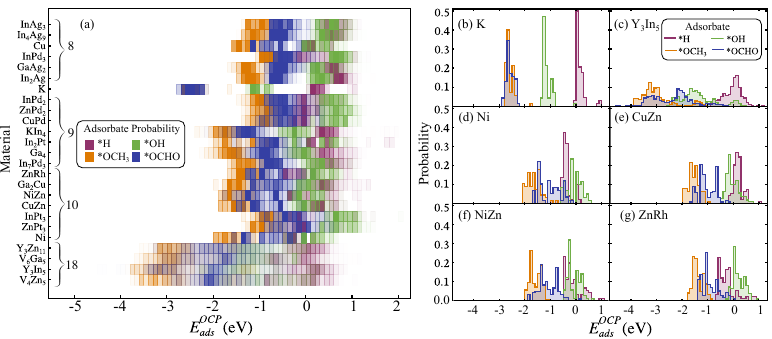}
    \caption{\textbf{AEDs for selected subset of materials:} (a) AEDs for 25 materials with the amplitude of the distribution encoded in the intensity of the corresponding color; detailed AEDs for  (b) K, (c) Y$_3$In$_5$, (d) Ni, (e) CuZn, (f) NiZn, (g) ZnRh.  The numbers on the \textit{y}-axis of panel (a) indicate the cluster numbers defined in Fig. \ref{fig_hierarchy}.}
    \label{fig_AED}
\end{figure}

Lastly, to compile the AEDs, we examined the relaxed configurations.
For many distinct initial configurations of identical adsorbates, materials, and facets that converged to the same final structure, only one of them is considered in the AED.
In our final compilation, we transformed all AEDs into histograms that depict the probability distribution of adsorption sites falling within 0.1~eV energy intervals.
Each AED was normalized, ensuring that the aggregate probability of adsorption sites per adsorbate and material equaled one.
This standardization facilitates direct comparisons across materials with different numbers of adsorption sites, which can range from several tens to nearly 10,000 for a single material, depending upon the complexity and symmetry of its bulk structure.
For illustrative purposes, Fig.~\ref{fig_AED} displays examples of AEDs for selected materials.
The AEDs for all investigated materials is shown in Fig.~S1 in the Supplementary Information~\ref{supple}. 

Inspection of Fig.~\ref{fig_AED} and Fig~S1 reveals that adsorption energies span a wide interval from -7.42~eV to 2.40~eV. We included energies above zero, although positive adsorption energies are typically indicative of molecular desorption. However, the adsorption energies reported in this work do not include entropy and pressure terms, which could shift the energies to more negative values. Secondly, the adsorption energy of radicals is somewhat ill-defined if different desorption channels are conceivable. Since our objective is to achieve a qualitative comparison across materials, the price energy zero is of no relevance, as long as it is chosen consistently.

The AEDs exhibit varying dispersion and forms, indicating fluctuations in adsorption energy and related activity levels across the material space. The adsorption energies of *OCH$_3$ ($E_{ads}$) are generally the lowest, followed by those of *OCHO, which are approximately 0.5--1 eV higher. However, certain materials, such as K (illustrated in Fig.~\ref{fig_AED}(b)), show unique distribution overlaps for *OCHO and *OCH$_3$. Meanwhile, *H and *OH have comparatively higher $E_{ads}$ values, although their order is inconsistent. For instance, in some cases, *H has the highest $E_{ads}$, particularly for K and Y$_{3}$In$_{5}$, whereas the opposite trend is observed for other materials like Ni. Single metal distributions are generally narrower and higher, as seen in the examples of K and Ni. Similarly, alloys composed of elements with high symmetry, such as CuZn, also exhibit narrow AEDs. 

If the AEDs of a material predominantly align around the adsorption energy linked to maximal activity according to the Sabatier principle, the material is a strong candidate for a good catalyst. Conversely, complex alloys with low symmetry, such as Y$_{3}$In$_{5}$ (shown in the lower section of Fig.~\ref{fig_AED}(a) and in Fig.~\ref{fig_AED}(c)), display broad AED spreads. Extremely low adsorption energies can lead to catalyst poisoning, while excessively high energies can significantly reduce catalytic activity. Therefore, broad distributions are less desirable, as only a small portion of the material’s surface contributes effectively to catalytic processes.

\subsubsection*{Unsupervised Learning: Catalyst Discovery}

\begin{figure}
	\centering 
	\includegraphics[width=0.6\textwidth]{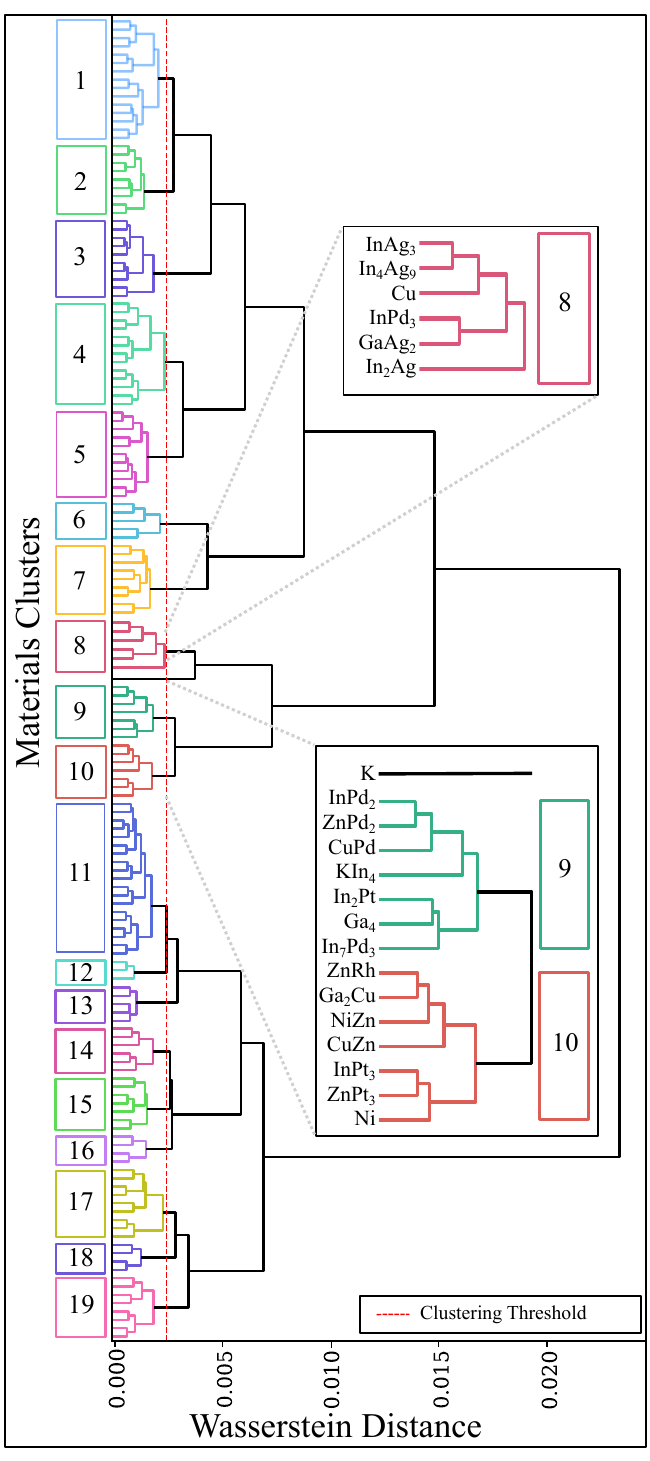}
    \caption{\textbf{Hierarchical clustering of the materials, based on the Wasserstein distances:}
The graph shows that all 159 materials were assembled into 19 clusters, based on the similarity between of the AEDs, with the exception of potassium, which is dissimilar to nearby materials and forms a single, non-numbered cluster. 
This can be seen in detail in the insets, where materials in clusters 8, 9 and 10 are shown. The cluster 10 contains several alloys, which are part of known high-yield catalysts as well as new potentially active materials.}
    \label{fig_hierarchy}
\end{figure}

Although the ideal AEDs for the four adsorbates remain unknown, it is feasible to approximate their reactivity using AEDs based on their resemblance to previously identified, efficacious catalytic materials.
In this context, our AEDs can be conceptualized as four-dimensional probability distributions.
To quantify similarities across AEDs of different materials, we employ the Wasserstein distance as the metric \cite{ramdas2017wasserstein}.
By computing Wasserstein distances for all possible material combinations, we construct a distance matrix.
To interpret the distance matrix, we apply hierarchical agglomerative clustering with Ward's linkage \cite{ward1963hierarchical}, which facilitates the identification of materials with similar AEDs. 
The outcomes of this clustering analysis are depicted in Fig.~\ref{fig_hierarchy}.

For a clustering threshold corresponding to a Wasserstein distance of $2.5 \times 10^{-3}$, we arrive at a total of 19 distinct clusters, with potassium (K) forming its own, isolated, unnumbered cluster.
The separation between clusters 11 to 19 and clusters 1 to 10 is considerable. The distinguishing feature is the broadness of the AEDs. The distributions in clusters 11 to 19 are noticeably broader than in clusters 1 to 10. Representative examples are depicted in Fig.~\ref{fig_AED}(a). The four materials at the bottom of the figure (Y$_3$Zn$_{11}$, V$_6$Ga$_5$, Y$_3$In$_5$, V$_4$Zn$_5$) pertain to cluster 18, whereas the rest belong to clusters 8 to 10.
Further details are available in Fig.~S1 of the Supplementary Information, presenting the clustering of all considered materials. 
AEDs exhibit variability across distinct clusters (1 -- 10) but show remarkable similarity within each individual cluster. For example, the AEDs for Ni, CuZn, NiZn, and ZnRh illustrated in Fig.~\ref{fig_AED}(d-g) belong to the same cluster.

Clusters 8 through 10 aggregate into a larger  cluster, from hereon denoted macro-cluster, with relatively homogeneous AEDs. It encompasses materials such as Cu, a notably active component within known Cu/ZnO/Al$_2$O$_3$ catalysts \cite{amann2022,bahri2022meta-analysis}. 
The clusters also contain non-Cu materials such as Zn-Pd, Pd-In, Pt-In, and Ni-Zn in different compositions that have been reported as catalytic converters of CO$_2$ to methanol~\cite{bahri2022meta-analysis, Dongapure_ChemCatChem23_NiZn}.
Also different compositions of the bimetallic alloys Ga-Ag, In-Ag, K-In, Zn-Rh, and Zn-Pt that, to our knowledge, have not been tested for CO$_2$ to methanol conversion are grouped with these materials.
While most of the materials in the macro-cluster
have either shown good catalytic performance or some have not been tested, potassium (K) (the lone non-numbered cluster), as a pure metal, is likely to undergo rapid oxidation under reaction conditions.
Therefore, we anticipate that the macro-cluster, consisting of clusters 8-10, is likely too diverse to pinpoint only catalytically active materials.

Upon closer inspection, ZnRh and ZnPt$_3$ stand out as new candidates. They are part of cluster 10, which also includes Ga$_2$Cu, NiZn, InPt$_3$, Ni and mainly CuZn, but have not been tested for CO$_2$ to methanol conversion. 
While the exact material composition of the Cu and ZnO-based catalysts during the proceeding reaction is still debated \cite{Beck2024, BehrensCu211Sci21, Laudenschleger2020}, some studies suggest that the formation of a Cu-Zn alloy enhances the activity. This Cu-Zn alloy, which is part of cluster 10, is believed to contribute to increased activity \cite{BehrensCu211Sci21, Studt2014GaNi, amann2022}. 
Similarly, NiZn has also been identified as an effective CO$_2$ catalyst \cite{Dongapure_ChemCatChem23_NiZn}. Catalysts such as Cu/Ga/ZnO \cite{CuGaTOYIR01}, Cu/Ga/SiO$_2$ \cite{Medina2017_GaCu}  and Pt/In$_2$O$_3$ \cite{PtInMen2019}, known for their high methanol yield, may include Ga$_2$Cu and InPt$_3$ alloys, respectively. Finally, Ni is often part of catalysts for CO$_2$ transformation to methane~\cite{Tang2024,HuNiBasedACS22}.
The strong catalytic activity of Ga$_2$Cu, NiZn, InPt$_3$, Ni and mainly CuZn in this cluster suggest that also ZnRh and ZnPt$_3$ should have a high activity.

 To conclude this section, we reiterate that our approach groups catalyst materials according to their computed AED similarity. To ascribe meaning to certain similarities we observe, we currently rely on experimental reports of the catalytically active materials. Our proposals for interesting candidates are based on the assumption that AED similarity with a known good catalyst is a meaningful indicator for promising catalytic activity. Since the  catalyst composition and microstructure are often not reported or not known, ``active sites'' or details of the catalytic mechanisms also remain opaque \cite{amann2022, lempelto2023, Laudenschleger2020, Beck2024}. In this context, our AED descriptor remains an attempt to find proxies for  complex processes. It goes beyond the common practice of focusing on single adsorption energies in ``active sites'', but could certainly be extended in future work and in collaboration with more detailed experimental investigations.

\subsubsection*{Statistical Analysis and Discussion:}

\begin{table}
\caption{\textbf{Extract of the statistical analysis on the AEDs:} The minimum of the OCP MLFF predicted adsorption energies $E_{\text{ads}}^{\text{min}}$ which is basically comparable to the adsorption energy used in Sabatier principle. We show the predicted $E_{\text{ads}}^{\text{min}}$ for the same materials as in Fig.~\ref{fig_AED} (i.e. clusters 8 -- 10 and 18), for all the considered adsorbates. The materials of cluster 10 are highlighted in bold.}
\label{tab:min_energies}
\begin{tabular}{lrrrr}
\toprule
Material & *H $E_{\text{ads}}^{\text{min}}$ & *OH $E_{\text{ads}}^{\text{min}}$ & *OCH$_3$ $E_{\text{ads}}^{\text{min}}$ & *OCHO $E_{\text{ads}}^{\text{min}}$ \\
\midrule
InAg$_{3}$ & 0.24 & 0.09 & -0.88 & -1.41 \\
In$_{4}$Ag$_{9}$ & 0.12 & -0.07 & -0.93 & -1.53 \\
Cu & -0.14 & -0.18 & -1.29 & -1.64 \\
InPd$_{3}$ & -0.54 & 0.27 & -1.04 & -1.22 \\
GaAg$_{2}$ & 0.22 & -0.25 & -1.06 & -1.80 \\
In$_{2}$Ag & 0.27 & -0.06 & -0.96 & -1.59 \\

\hline
K & -0.01 & -1.31 & -2.81 & -2.77 \\
\hline
InPd$_{2}$ & -0.39 & -0.04 & -1.09 & -1.57 \\
ZnPd$_{2}$ & -0.38 & 0.06 & -1.11 & -1.44 \\
CuPd & -0.44 & 0.03 & -1.12 & -1.41 \\
KIn$_{4}$ & -0.40 & -1.11 & -2.36 & -2.46 \\
In$_{2}$Pt & -0.33 & -0.54 & -1.69 & -2.12 \\
Ga$_{4}$ &  -0.71 & -0.98 & -1.73 & -2.36 \\
In$_{7}$Pd$_{3}$ & -0.36 & -0.33 & -1.20 & -1.80 \\
\hline
\textbf{ZnRh} & -0.63 & -0.30 & -1.39 & -1.71 \\
\textbf{Ga$_{2}$Cu} & -0.20 & -0.92 & -1.36 & -2.06 \\
\textbf{NiZn} & -0.49 & -0.49 & -1.81 & -1.93 \\
\textbf{CuZn} & -0.19 & -0.44 & -1.67 & -1.94 \\
\textbf{InPt$_{3}$} & -0.59 & 0.13 & -1.09 & -1.41 \\
\textbf{ZnPt$_{3}$} & -0.43 & 0.10 & -0.98 & -1.40 \\
\textbf{Ni}& -0.53 & -0.54 & -1.66 & -2.05 \\
\hline
Y$_{3}$Zn$_{11}$ & -0.89 & -2.22 & -3.92 & -3.68 \\
V$_{6}$Ga$_{5}$ & -1.30 & -3.37 & -5.44 & -3.78 \\
Y$_{3}$In$_{5}$ & -0.88 & -2.51 & -4.61 & -3.93 \\
V$_{4}$Zn$_{5}$ & -1.13 & -2.29 & -3.72 & -3.87 \\
\bottomrule
\end{tabular}
\end{table}

AEDs could serve as a descriptor of activity, however, the vast number of parameters (at least 388 bins in the distribution) makes it challenging to analyze them manually. 
To further our insight into the generated data, we conducted a statistical analysis of AEDs (SAAEDs) that facilitates comparison with previous adsorption energy-based studies. 
An example can be seen in Table~\ref{tab:min_energies}, where we present the minimum adsorption energies for a subset of materials featured in Fig~\ref{fig_AED}(a).

Our SAAED analysis revisits individual binding energies and connects to the Sabatier principle. 
For instance, the results for *OH, *OCHO, and *OCH$_3$ can be compared to the volcano plot in Studt \textit{et al.} \cite{Studt2014GaNi}, that relates the catalytic activity of the studied materials to the oxygen adsorption energy. In line with our approach, their work compares potential catalyst materials to Cu, although their focus lies on single-facet surfaces. Following previous findings that the Cu(211) facet is more active than the close-packed Cu(111) surface~\cite{BehrensCu211Sci21}, Studt \textit{et al.} use Cu(211) as their reference.
The catalytic activity of Cu is further enhanced when Zn is added to the Cu(211) surface (referred to as Cu+Zn in the article). The oxygen adsorption energy decreases upon Zn addition, which indicates that the optimal oxygen adsorption energy should be lower than its minimal adsorption energy on the Cu(211) surface.
Our data is consistent with those findings for the majority of our promising candidate materials.
The minimal adsorption energy ($E_{ads}^{min}$) for all the oxygen-containing adsorbates on the majority of the materials in cluster 10 (highlighted in Table~\ref{tab:min_energies}), including ZnRh,
lies below that of Cu (our Cu data also covers the (211) surface) and is closely aligned across the materials.
The exceptions are InPt$_3$ and ZnPt$_3$, in which the minima lie slightly above those of Cu, while both materials exhibit similar $E_{ads}^{min}$ for all other adsorbates.
This difference suggests that InPt$_3$ and ZnPt$_3$ may feature slightly different CO$_2$ conversion mechanisms.

Using minimum adsorption energies derived from ML models is comparable to previously studied methods for identifying global minima~\cite{Lan2023, Chen2024}.
Although the techniques by Lan \textit{et al.}~\cite{Lan2023} and Chen \textit{et al.}~\cite{Chen2024} might be more appropriate for the straightforward application of the Sabatier principle, our approach excels in providing more comprehensive information on various facets of catalytic materials.
We have compiled this information for selected materials in Table~S1 in the Supplementary Information.
For example, the AED spread across energies, which can be deduced from the standard deviation $E_{\text{ads}}^{\text{std}}$, provides information about the percentage of the surface area usable for catalytic conversion.

Ultimately, both AEDs and SAAEDs, available on Zenodo~\cite{dataset}, can serve as material fingerprints.
The SAAED acts like a materials descriptor, similar to the Magpie descriptor~\cite{Ward2016}, but can be adapted to specific reactions through the choice of adsorbates, offering more detailed and relevant material information.
Optionally, specific descriptors (AED, SAAED) and general descriptors (like Magpie \cite{Ward2016}) may be combined to enhance the information that might be lacking in ML models from theoretical calculations.

Both catalyst descriptors are tailored  to perform extensive searches for catalytically active candidates. They do not, however, include effects of the support, additives, preparation procedures and operando states that could change the morphology of the catalyst (e.g., nanoparticle sizes or areas of different facets). Our proposed AED descriptor does not take the facet area into account, and is therefore insensitive to morphology changes of the catalysts under reaction conditions. 
In principle, a Wulff construction could introduce better facet information. 
However, it also cannot account for support effects, additives or preparation conditions.

Our methodology facilitates high-throughput screening of metallic catalyst candidates. At present, effects of co-operating oxides such as ZnO, In$_2$O$_3$, and ZrO$_2$ that have been observed experimentally, are not considered. Such oxides affect the electronic structure and adsorption energy landscape at the metal-oxide interface~\cite{Wang2019_MaZrOx, Cheula2024_DopantsZirconia, lempelto2023, Cannizzaro2023, Beck2024} and should be included in future versions of our descriptor.
For instance, incorporating general descriptors (e.g., Magpie) for co-catalysts and support materials could provide additional information to decide which active material-support combination should be investigated further in experimental testing.

Additionally, the choice of adsorbates can also influence the effectiveness of the AED descriptor. Our study focuses on four most relevant intermediates, observed on Cu(211) surface~\cite{amann2022}. Studies on different materials, such as Ni-ZrO$_2$~\cite{Cannizzaro2023, Gao2022}, suggest that other intermediates or by-products like CO could play an important role in the hydrogenation mechanism. Thus,
further investigations could extend the set of adsorbates to better capture various reaction paths and therefore material-specific activity.

Our workflow clusters materials with high CO$_2$ conversion efficiency, but the materials can vary in their selectivity towards methanol or methane \cite{CuGaTOYIR01, bahri2022meta-analysis, PtInMen2019, Dongapure_ChemCatChem23_NiZn, LIU2024NiZr}. As the reaction conditions, preparation procedures, or the interaction with support materials seem to affect the selectivity \cite{Dongapure_ChemCatChem23_NiZn, LIU2024NiZr}, our proposed catalyst candidates should thus be tested under various conditions to investigate their optimal selectivity towards methanol.

To finalize the analysis of our results, the similarity of the SAAED and Wasserstein distances of ZnRh and ZnPt$_3$ to good catalysts in the literature suggests that they could be potential catalyst candidates.
As Cu-based catalysts are known for their vulnerability to degradation~\cite{ganesh2014renewable}, it is therefore reasonable to pre-examine these materials also in terms of stability. 
Given the harsh reaction conditions, mainly temperatures around 800~K \cite{bahri2022meta-analysis}, the melting temperature of the catalyst is directly related to the stability of the catalyst. 
The melting temperature of both ZnRh and ZnPt$_3$  is higher than that of pure copper or CuZn~\cite{matproj2013}, suggesting that our new candidates could also be more durable.

\subsubsection*{Summary:}

In summary, we have established a fast and reliable computational approach for discovering new catalyst candidates for the conversion of CO$_2$ to methanol utilizing data-driven methodologies such as MLFFs and hierarchical clustering. 
Beginning with a list of potential metallic elements, we extracted experimentally verified materials from the Materials Project database.
By integrating tools from \texttt{fairchem}, mainly OCP MLFFs, we created an extensive database of adsorption energies for a wide range of materials facets and possible adsorption sites.
We compiled this information to obtain a novel material descriptor, AED, which offers a more effective representation of the complex nature of heterocatalysts compared to standard methods.
By carefully choosing the adsorbates, the descriptor can be tailored to provide the most information for any heterocatalytic reaction under study.
Through efficient sampling for validation, we were able to quantify the quality of our workflow with a minimal number of DFT calculations while ensuring the high quality of our database.
We grouped the materials by their AED similarity using statistical methods and clustering.
This allowed us to pinpoint promising new candidates, namely ZnRh and ZnPt$_3$, based on their resemblance to known effective catalysts.
Our results indicate that AEDs, together with statistical analysis, can serve as material fingerprints, aiding in the prediction of catalyst activity and accelerating the discovery process. 

\section{Methods}\label{methods}

\subsection{Bulk Preparation}\label{bulkprep}
We sourced the bulk geometries of experimentally observed metals and alloys of select elements from the Materials Project Database \cite{matproj2013}. These structures were selected on the basis of their stability, as indicated by cohesive energies located on the convex hull. We optimized the bulk geometries using the RPBE functional \cite{PhysRevB.59.7413} as implemented in the Vienna ab-initio Simulation Package (\texttt{VASP}) \cite{vasp1_PhysRevB.49.14251, vasp2_PhysRevB.54.11169}. A plane-wave cutoff of 500~eV was used in the first attempt to relax the structures. If the initial run did not converge, we increased the cut-off to 550~eV. We sampled the Brillouin zone with a \textit{k}-point spacing of 0.17~\AA{}$^{-1}$. The calculations were automated using the workflows developed by Atomate \cite{atomate_2017} based on the Pymatgen \cite{pymatgen_2013}, Custodian \cite{custodian_2013} and Fireworks \cite{firworks_2015} libraries. 

\subsection{Surface Generation and Selection}\label{slabgen}

We generated all symmetrically distinct surfaces of Miller indices $\in \{-2,-1,0,...,2\}$ from the relaxed bulk structures using the workflow implemented by OCP, i.e. \texttt{fairchem}  \cite{chanussot2021open, fairchem}. 
We fixed the thickness of the slabs to 7~\AA{} with a vacuum of 20~\AA{} along the \textit{z}-direction. These parameters were chosen to be consistent with the OC20 dataset that was used to train the OCP MLFFs to maintain high prediction accuracy~\cite{chanussot2021open}. 
With the pre-trained \texttt{gemnet-oc} MLFF~\cite{gasteiger2022gemnetocdevelopinggraphneural}, we relaxed the surfaces and obtain their \textit{total energy}.
When encountering different surface terminations for a given facet, i.e., different absolute positions of the surface plane, we retained only the structure with the lowest energy. 

\subsection{Generation and Relaxation of Adsorbate-Surface Configurations}\label{adslabs}

The selected surfaces were used to generate adsorbate-surface configurations using neutral fragments of *H, *OH, *OCHO and *OCH$_3$ by means of the \texttt{fairchem} input generation workflow~\cite{fairchem}. The systems were relaxed using an \textit{adsorption energy}-based OCP \texttt{equiformer\_V2} equivariant graph neural network MLFF to obtain the relaxed geometries. 
The corresponding adsorption energy $E_{ads}$ follows the OCP convention  $E_{ads} = E_{system} - E_{surface} - E_{adsorbate}$, where the gas phase energy  $E_{adsorbate}$ is 
 in this case the energy of radicals, and therefore hard to define. 
The OCP MLFFs are using the definition of  
Chanussot \textit{et. al.}~\cite{chanussot2021open}
, which is a linear combination of atomic energies for H, O and C, derived from H$_2$, H$_2$O and CO, respectively.
Here, $E_{adsorbate}$ is --3.477 eV for *H, --10.552 eV for *OH, --24.917 eV for *OCH$_3$ and --25.161 eV for *OCHO.
As our work is focusing on identifying AED similarities or differences rather than specific adsorbate values, different adsorbate references would merely shift all AEDs by the same value, but will not affect our results.
The force convergence criterion for the adsorbate relaxations was set to 0.03~eV/\AA{}.
The \texttt{equiformer\_V2} relaxations were performed using \texttt{NVIDIA Ampere A100} GPUs, completing each relaxation in just a few seconds, whereas corresponding DFT calculations would require several hours on two \texttt{AMD Rome 7H12} CPUs with 64 cores each. Relaxed configurations of the same material and facet were compared to identify redundancies. If two or more configurations exhibited identical adsorbate geometries (and thus energies) within a tolerance of 0.1~\AA{} in each spatial direction, only one configuration was retained for further consideration. Nonunique configurations and their corresponding adsorption energies were excluded. 
With this we ensured that the remaining data represents meaningful adsorption positions and thus capture the material's true energetic landscape.

We aggregated the adsorption energies into AEDs for every adsorbate and material. The AED is represented by a histogram of 0.1~eV wide bins, that are centered along energies $\{-7.4, -7.3, .... 2.4\}$. The histogram settings were chosen so that each AED has at least five empty bins on either side, which is important for further analysis. We normalized the AEDs by dividing the entry in each bin by the number of unique adsorption configurations.

\subsection{Estimation of the Mean Absolute Error for the Adsorption Energy Prediction}\label{validation}
We initially evaluated the performance of the \texttt{equiformer\_V2} model on Pt, Zn, and NiZn. 
For this evaluation, we used the adsorbate-surface geometries relaxed with \texttt{equiformer\_V2} and the clean surface geometries optimized using \texttt{gemnet-oc}.
We then computed the DFT adsorption energies without further relaxations. We used the same \texttt{VASP} settings as for the generation of the OC20 dataset \cite{chanussot2021open}, but increased the plane-wave cutoff to 450~eV. This adjustment was made on the basis of our preliminary estimate for selected materials to ensure high accuracy of both predicted and calculated adsorption energies. The resulting adsorption energies are shown in Figure~\ref{fig_MAE}. 

To determine the EMAE, we employed a computationally efficient strategy to validate the predictions of the OCP MLFF model.
From the AED of a given material-adsorbate pair, we chose the three configurations that correspond to the mean, median, and maximum adsorption energy.
Subsequently, we carried out single-point DFT calculations for the corresponding adsorbate-surface systems and clean surfaces to compute the adsorption energies.
 
The EMAE was determined as the average of the three absolute errors.
As discussed in Section~\ref{resdis}, materials with EMAEs exceeding 0.25~eV were excluded from further analysis due to their insufficient accuracy.

\subsection{Unsupervised Learning}\label{unsup}

We concatenated the four AEDs for each material into a single distribution, where the added buffer (see Section~\ref{adslabs}) ensures no possible overlap between distributions. We then computed Wasserstein distances for all pairs of materials to create a single distance matrix. Given two 1D probability mass functions, $\mu$ and $\nu$, the first Wasserstein distance between the distributions is defined as~\cite{2020SciPy-NMeth}:

\begin{equation}
W_1(\mu, \nu) = \inf_{\gamma \in \Gamma(\mu, \nu)} \int_{\mathbb{R} \times \mathbb{R}} ||x-y|| d\gamma(x,y)
\end{equation}
where $\Gamma(\mu, \nu)$ is the set of (probability) distributions on $\mathbb{R} \times \mathbb{R}$ whose marginals are $\mu$ and $\nu$ on the first and second factors, respectively. For a given value $x$, $\mu(x)$ gives the probability of $\mu$ at position $x$, and the same for $\nu(x)$.

Using these Wasserstein distances, we performed agglomerative hierarchical clustering with Ward linkage \cite{müllner2011modernhierarchicalagglomerativeclustering, bar2001hierarchical, ward1963hierarchical}. We utilized the Python-based SciPy library \cite{2020SciPy-NMeth} to compute the distances and to perform the clustering on the distance matrix. 
We used a clustering threshold of 0.0025 to define the maximum distance at which clusters are merged.



\backmatter

\bmhead{Supplementary information}\label{supple}

The Supplementary  information contains:~\\
Table~S1 -- All 158 materials used in final evaluation;
Table~S2 -- All materials removed from the final evaluation;
Figure~S1 -- Adsorption energy distributions and hierarchical clustering of all 158 materials;
Table~S3 -- Statistical analysis on the AEDs for a subset of materials shown in Fig.~\ref{fig_AED}.

\section*{Declarations}

\bmhead{Data Availability:}
The data published in this study can be found in \cite{dataset}.

\bmhead{Acknowledgements}
The authors would like to thank Annukka Santasalo-Aarnio and Arpad Toldy for fruitful discussions.
O.K. and P.P. express their gratitude to Kirby Broderick, Adeesh Koluru, Brook Wander, John Kitchin and other Kitchin research group members at Carnegie Mellon University and Zachary Ulissi at Meta, for their help with the OCP models.
This project received funding from the European Union – NextGenerationEU instrument and the Research Council of Finland's AICon project (grant number no. 348179). 
The authors gratefully acknowledge CSC – IT Center for Science, Finland, and the Aalto Science-IT project for the generous computational resources.

\bmhead{Author contributions}
P.P. and O.K. created the workflow and proceeded the calculations.
P.R. supervised the work.
All authors contributed to the manuscript. 

\bmhead{Competing Interests}
The authors declare no conflicts of interest.


\noindent

\bigskip
\begin{flushleft}%
Editorial Policies for:

\bigskip\noindent
Springer journals and proceedings: \url{https://www.springer.com/gp/editorial-policies}

\bigskip\noindent
Nature Portfolio journals: \url{https://www.nature.com/nature-research/editorial-policies}

\bigskip\noindent
\textit{Scientific Reports}: \url{https://www.nature.com/srep/journal-policies/editorial-policies}

\bigskip\noindent
BMC journals: \url{https://www.biomedcentral.com/getpublished/editorial-policies}
\end{flushleft}



\bibliography{sn-bibliography}


\begin{thebibliography}{70}
\ifx \bisbn   \undefined \def \bisbn  #1{ISBN #1}\fi
\ifx \binits  \undefined \def \binits#1{#1}\fi
\ifx \bauthor  \undefined \def \bauthor#1{#1}\fi
\ifx \batitle  \undefined \def \batitle#1{#1}\fi
\ifx \bjtitle  \undefined \def \bjtitle#1{#1}\fi
\ifx \bvolume  \undefined \def \bvolume#1{\textbf{#1}}\fi
\ifx \byear  \undefined \def \byear#1{#1}\fi
\ifx \bissue  \undefined \def \bissue#1{#1}\fi
\ifx \bfpage  \undefined \def \bfpage#1{#1}\fi
\ifx \blpage  \undefined \def \blpage #1{#1}\fi
\ifx \burl  \undefined \def \burl#1{\textsf{#1}}\fi
\ifx \doiurl  \undefined \def \doiurl#1{\url{https://doi.org/#1}}\fi
\ifx \betal  \undefined \def \betal{\textit{et al.}}\fi
\ifx \binstitute  \undefined \def \binstitute#1{#1}\fi
\ifx \binstitutionaled  \undefined \def \binstitutionaled#1{#1}\fi
\ifx \bctitle  \undefined \def \bctitle#1{#1}\fi
\ifx \beditor  \undefined \def \beditor#1{#1}\fi
\ifx \bpublisher  \undefined \def \bpublisher#1{#1}\fi
\ifx \bbtitle  \undefined \def \bbtitle#1{#1}\fi
\ifx \bedition  \undefined \def \bedition#1{#1}\fi
\ifx \bseriesno  \undefined \def \bseriesno#1{#1}\fi
\ifx \blocation  \undefined \def \blocation#1{#1}\fi
\ifx \bsertitle  \undefined \def \bsertitle#1{#1}\fi
\ifx \bsnm \undefined \def \bsnm#1{#1}\fi
\ifx \bsuffix \undefined \def \bsuffix#1{#1}\fi
\ifx \bparticle \undefined \def \bparticle#1{#1}\fi
\ifx \barticle \undefined \def \barticle#1{#1}\fi
\bibcommenthead
\ifx \bconfdate \undefined \def \bconfdate #1{#1}\fi
\ifx \botherref \undefined \def \botherref #1{#1}\fi
\ifx \url \undefined \def \url#1{\textsf{#1}}\fi
\ifx \bchapter \undefined \def \bchapter#1{#1}\fi
\ifx \bbook \undefined \def \bbook#1{#1}\fi
\ifx \bcomment \undefined \def \bcomment#1{#1}\fi
\ifx \oauthor \undefined \def \oauthor#1{#1}\fi
\ifx \citeauthoryear \undefined \def \citeauthoryear#1{#1}\fi
\ifx \endbibitem  \undefined \def \endbibitem {}\fi
\ifx \bconflocation  \undefined \def \bconflocation#1{#1}\fi
\ifx \arxivurl  \undefined \def \arxivurl#1{\textsf{#1}}\fi
\csname PreBibitemsHook\endcsname

\bibitem[\protect\citeauthoryear{Ye et~al.}{2019}]{ye2019co2}
\begin{botherref}
\oauthor{\bsnm{Ye}, \binits{R.-P.}},
\oauthor{\bsnm{Ding}, \binits{J.}},
\oauthor{\bsnm{Gong}, \binits{W.}},
\oauthor{\bsnm{Argyle}, \binits{M.D.}},
\oauthor{\bsnm{Zhong}, \binits{Q.}},
\oauthor{\bsnm{Wang}, \binits{Y.}},
\oauthor{\bsnm{Russell}, \binits{C.K.}},
\oauthor{\bsnm{Xu}, \binits{Z.}},
\oauthor{\bsnm{Russell}, \binits{A.G.}},
\oauthor{\bsnm{Li}, \binits{Q.}},
\oauthor{\bsnm{Fan}, \binits{M.}},
\oauthor{\bsnm{Yao}, \binits{Y.-G.}}:
{CO$_2$} hydrogenation to high-value products via heterogeneous catalysis.
Nature Communications
\textbf{10}(5698)
(2019)
\doiurl{10.1038/s41467-019-13638-9}
\end{botherref}
\endbibitem

\bibitem[\protect\citeauthoryear{Rohr et~al.}{2019}]{ROHR201933}
\begin{barticle}
\bauthor{\bsnm{Rohr}, \binits{B.A.}},
\bauthor{\bsnm{Singh}, \binits{A.R.}},
\bauthor{\bsnm{N\o{}rskov}, \binits{J.K.}}:
\batitle{A theoretical explanation of the effect of oxygen poisoning on industrial haber-bosch catalysts}.
\bjtitle{Journal of Catalysis}
\bvolume{372},
\bfpage{33}--\blpage{38}
(\byear{2019})
\doiurl{10.1016/j.jcat.2019.01.042}
\end{barticle}
\endbibitem

\bibitem[\protect\citeauthoryear{Saito et~al.}{1996}]{96CO2_methanol}
\begin{barticle}
\bauthor{\bsnm{Saito}, \binits{M.}},
\bauthor{\bsnm{Fujitani}, \binits{T.}},
\bauthor{\bsnm{Takeuchi}, \binits{M.}},
\bauthor{\bsnm{Watanabe}, \binits{T.}}:
\batitle{Development of copper\/zinc oxide-based multicomponent catalysts for methanol synthesis from carbon dioxide and hydrogen}.
\bjtitle{Applied Catalysis A: General}
\bvolume{138}(\bissue{2}),
\bfpage{311}--\blpage{318}
(\byear{1996})
\doiurl{10.1016/0926-860X(95)00305-3}
\end{barticle}
\endbibitem

\bibitem[\protect\citeauthoryear{Nyári et~al.}{2020}]{nyari2020techno}
\begin{barticle}
\bauthor{\bsnm{Nyári}, \binits{J.}},
\bauthor{\bsnm{Magdeldin}, \binits{M.}},
\bauthor{\bsnm{Larmi}, \binits{M.}},
\bauthor{\bsnm{Järvinen}, \binits{M.}},
\bauthor{\bsnm{Santasalo-Aarnio}, \binits{A.}}:
\batitle{Techno-economic barriers of an industrial-scale methanol {CCU}-plant}.
\bjtitle{Journal of {CO$_2$} Utilization}
\bvolume{39},
\bfpage{101166}
(\byear{2020})
\doiurl{10.1016/j.jcou.2020.101166}
\end{barticle}
\endbibitem

\bibitem[\protect\citeauthoryear{Ganesh}{2014}]{ganesh2014renewable}
\begin{barticle}
\bauthor{\bsnm{Ganesh}, \binits{I.}}:
\batitle{Conversion of carbon dioxide into methanol – a potential liquid fuel: Fundamental challenges and opportunities (a review)}.
\bjtitle{Renewable and Sustainable Energy Reviews}
\bvolume{31},
\bfpage{221}--\blpage{257}
(\byear{2014})
\doiurl{10.1016/j.rser.2013.11.045}
\end{barticle}
\endbibitem

\bibitem[\protect\citeauthoryear{Tran and Ulissi}{2018}]{Tran2018el-chem}
\begin{barticle}
\bauthor{\bsnm{Tran}, \binits{K.}},
\bauthor{\bsnm{Ulissi}, \binits{Z.W.}}:
\batitle{Active learning across intermetallics to guide discovery of electrocatalysts for {CO$_2$} reduction and {H$_2$} evolution}.
\bjtitle{Nature Catalysis}
\bvolume{1},
\bfpage{696}--\blpage{703}
(\byear{2018})
\doiurl{10.1038/s41929-018-0142-1}
\end{barticle}
\endbibitem

\bibitem[\protect\citeauthoryear{Wang et~al.}{2022}]{wang2022co2}
\begin{barticle}
\bauthor{\bsnm{Wang}, \binits{L.}},
\bauthor{\bsnm{Etim}, \binits{U.J.}},
\bauthor{\bsnm{Zhang}, \binits{C.}},
\bauthor{\bsnm{Amirav}, \binits{L.}},
\bauthor{\bsnm{Zhong}, \binits{Z.}}:
\batitle{{CO$_2$} activation and hydrogenation on {Cu-ZnO/Al$_2$O$_3$} nanorod catalysts: An in situ {FTIR} study}.
\bjtitle{Nanomaterials}
\bvolume{12}(\bissue{15}),
\bfpage{2527}
(\byear{2022})
\doiurl{10.3390/nano12152527}
\end{barticle}
\endbibitem

\bibitem[\protect\citeauthoryear{Li et~al.}{2023}]{li2023deactivation}
\begin{barticle}
\bauthor{\bsnm{Li}, \binits{D.}},
\bauthor{\bsnm{Wang}, \binits{Z.}},
\bauthor{\bsnm{Jin}, \binits{S.}},
\bauthor{\bsnm{Zhu}, \binits{M.}}:
\batitle{Deactivation and regeneration of the commercial {Cu/ZnO/Al$_2$O$_3$} catalyst in low-temperature methanol steam reforming}.
\bjtitle{Sci China Chem}
\bvolume{66},
\bfpage{3645}--\blpage{3652}
(\byear{2023})
\doiurl{10.1007/s11426-023-1789-3}
\end{barticle}
\endbibitem

\bibitem[\protect\citeauthoryear{Bruix et~al.}{2019}]{Bruix2019}
\begin{barticle}
\bauthor{\bsnm{Bruix}, \binits{A.}},
\bauthor{\bsnm{Margraf}, \binits{J.T.}},
\bauthor{\bsnm{Andersen}, \binits{M.}},
\bauthor{\bsnm{Reuter}, \binits{K.}}:
\batitle{First-principles-based multiscale modelling of heterogeneous catalysis}.
\bjtitle{Nature Catalysis}
\bvolume{2},
\bfpage{659}--\blpage{670}
(\byear{2019})
\doiurl{10.1038/s41929-019-0298-3}
\end{barticle}
\endbibitem

\bibitem[\protect\citeauthoryear{Posada-Borbón and Grönbeck}{2021}]{posada2021firstprinciples}
\begin{barticle}
\bauthor{\bsnm{Posada-Borbón}, \binits{A.}},
\bauthor{\bsnm{Grönbeck}, \binits{H.}}:
\batitle{A first-principles-based microkinetic study of {CO$_2$} reduction to {CH$_3$OH} over {In$_2$O$_3$(110)}}.
\bjtitle{ACS Catal.}
\bvolume{11}(\bissue{15}),
\bfpage{9996}--\blpage{10006}
(\byear{2021})
\doiurl{10.1021/acscatal.1c01707}
\end{barticle}
\endbibitem

\bibitem[\protect\citeauthoryear{Che}{2013}]{CHE2013162}
\begin{barticle}
\bauthor{\bsnm{Che}, \binits{M.}}:
\batitle{Nobel prize in chemistry 1912 to {S}abatier: Organic chemistry or catalysis?}
\bjtitle{Catalysis Today}
\bvolume{218-219},
\bfpage{162}--\blpage{171}
(\byear{2013})
\doiurl{10.1016/j.cattod.2013.07.006} .
\bcomment{Catalysis: From the active sites to the processes}
\end{barticle}
\endbibitem

\bibitem[\protect\citeauthoryear{Medford et~al.}{2015}]{medford2015sabatier}
\begin{barticle}
\bauthor{\bsnm{Medford}, \binits{A.J.}},
\bauthor{\bsnm{Vojvodic}, \binits{A.}},
\bauthor{\bsnm{Hummelshøj}, \binits{J.S.}},
\bauthor{\bsnm{Voss}, \binits{J.}},
\bauthor{\bsnm{Abild-Pedersen}, \binits{F.}},
\bauthor{\bsnm{Studt}, \binits{F.}},
\bauthor{\bsnm{Bligaard}, \binits{T.}},
\bauthor{\bsnm{Nilsson}, \binits{A.}},
\bauthor{\bsnm{Nørskov}, \binits{J.K.}}:
\batitle{From the {S}abatier principle to a predictive theory of transition-metal heterogeneous catalysis}.
\bjtitle{Journal of Catalysis}
\bvolume{328},
\bfpage{36}--\blpage{42}
(\byear{2015})
\doiurl{10.1016/j.jcat.2014.12.033}
\end{barticle}
\endbibitem

\bibitem[\protect\citeauthoryear{Chen et~al.}{2024}]{Chen2024}
\begin{barticle}
\bauthor{\bsnm{Chen}, \binits{Z.W.}},
\bauthor{\bsnm{Li}, \binits{J.}},
\bauthor{\bsnm{Ou}, \binits{P.}},
\bauthor{\bsnm{Huang}, \binits{J.E.}},
\bauthor{\bsnm{Wen}, \binits{Z.}},
\bauthor{\bsnm{Chen}, \binits{L.}},
\bauthor{\bsnm{Yao}, \binits{X.}},
\bauthor{\bsnm{Cai}, \binits{G.}},
\bauthor{\bsnm{Yang}, \binits{C.C.}},
\bauthor{\bsnm{Singh}, \binits{C.V.}},
\bauthor{\bsnm{Jiang}, \binits{Q.}}:
\batitle{Unusual {S}abatier principle on high entropy alloy catalysts for hydrogen evolution reactions}.
\bjtitle{Nature Communications}
\bvolume{15}(\bissue{1}),
\bfpage{359}
(\byear{2024})
\doiurl{10.1038/s41467-023-44261-4}
\end{barticle}
\endbibitem

\bibitem[\protect\citeauthoryear{Jones et~al.}{2008}]{Jones_2008}
\begin{barticle}
\bauthor{\bsnm{Jones}, \binits{G.}},
\bauthor{\bsnm{Bligaard}, \binits{T.}},
\bauthor{\bsnm{Abild-Pedersen}, \binits{F.}},
\bauthor{\bsnm{Nørskov}, \binits{J.K.}}:
\batitle{Using scaling relations to understand trends in the catalytic activity of transition metals}.
\bjtitle{Journal of Physics: Condensed Matter}
\bvolume{20}(\bissue{6}),
\bfpage{064239}
(\byear{2008})
\doiurl{10.1088/0953-8984/20/6/064239}
\end{barticle}
\endbibitem

\bibitem[\protect\citeauthoryear{Vogt and Weckhuysen}{2022}]{vogt2022}
\begin{barticle}
\bauthor{\bsnm{Vogt}, \binits{C.}},
\bauthor{\bsnm{Weckhuysen}, \binits{B.M.}}:
\batitle{The concept of active site in heterogeneous catalysis}.
\bjtitle{Nature Reviews Chemistry}
\bvolume{6}(\bissue{2}),
\bfpage{89}--\blpage{111}
(\byear{2022})
\doiurl{10.1038/s41570-021-00340-y}
\end{barticle}
\endbibitem

\bibitem[\protect\citeauthoryear{Ward et~al.}{2016}]{Ward2016}
\begin{barticle}
\bauthor{\bsnm{Ward}, \binits{L.}},
\bauthor{\bsnm{Agrawal}, \binits{A.}},
\bauthor{\bsnm{Choudhary}, \binits{A.}},
\bauthor{\bsnm{Wolverton}, \binits{C.}}:
\batitle{A general-purpose machine learning framework for predicting properties of inorganic materials}.
\bjtitle{npj Computational Materials}
\bvolume{2}(\bissue{1}),
\bfpage{16028}
(\byear{2016})
\doiurl{10.1038/npjcompumats.2016.28}
\end{barticle}
\endbibitem

\bibitem[\protect\citeauthoryear{Ulissi et~al.}{2017}]{Ulissi2017}
\begin{barticle}
\bauthor{\bsnm{Ulissi}, \binits{Z.W.}},
\bauthor{\bsnm{Tang}, \binits{M.T.}},
\bauthor{\bsnm{Xiao}, \binits{J.}},
\bauthor{\bsnm{Liu}, \binits{X.}},
\bauthor{\bsnm{Torelli}, \binits{D.A.}},
\bauthor{\bsnm{Karamad}, \binits{M.}},
\bauthor{\bsnm{Cummins}, \binits{K.}},
\bauthor{\bsnm{Hahn}, \binits{C.}},
\bauthor{\bsnm{Lewis}, \binits{N.S.}},
\bauthor{\bsnm{Jaramillo}, \binits{T.F.}},
\bauthor{\bsnm{Chan}, \binits{K.}},
\bauthor{\bsnm{Nørskov}, \binits{J.K.}}:
\batitle{Machine-learning methods enable exhaustive searches for active bimetallic facets and reveal active site motifs for {CO$_2$} reduction}.
\bjtitle{ACS Catalysis}
\bvolume{7}(\bissue{10}),
\bfpage{6600}--\blpage{6608}
(\byear{2017})
\doiurl{10.1021/acscatal.7b01648}
\end{barticle}
\endbibitem

\bibitem[\protect\citeauthoryear{Andersen et~al.}{2019}]{Andersen2019}
\begin{barticle}
\bauthor{\bsnm{Andersen}, \binits{M.}},
\bauthor{\bsnm{Levchenko}, \binits{S.V.}},
\bauthor{\bsnm{Scheffler}, \binits{M.}},
\bauthor{\bsnm{Reuter}, \binits{K.}}:
\batitle{Beyond scaling relations for the description of catalytic materials}.
\bjtitle{ACS Catalysis}
\bvolume{9},
\bfpage{2752}--\blpage{2759}
(\byear{2019})
\doiurl{10.1021/acscatal.8b04478}
\end{barticle}
\endbibitem

\bibitem[\protect\citeauthoryear{Himanen et~al.}{2019}]{Himanen/Geurts/Foster/Rinke:2019}
\begin{barticle}
\bauthor{\bsnm{Himanen}, \binits{L.}},
\bauthor{\bsnm{Geurts}, \binits{A.}},
\bauthor{\bsnm{Foster}, \binits{A.S.}},
\bauthor{\bsnm{Rinke}, \binits{P.}}:
\batitle{Data-driven materials science: Status, challenges, and perspectives}.
\bjtitle{Advanced Science}
\bvolume{6}(\bissue{21}),
\bfpage{1900808}
(\byear{2019})
\doiurl{10.1002/advs.201900808}
{\href{https://arxiv.org/abs/https://advanced.onlinelibrary.wiley.com/doi/pdf/10.1002/advs.201900808}{{https://advanced.onlinelibrary.wiley.com/doi/pdf/10.1002/advs.201900808}}}
\end{barticle}
\endbibitem

\bibitem[\protect\citeauthoryear{Zhang et~al.}{2021}]{zhang2021}
\begin{barticle}
\bauthor{\bsnm{Zhang}, \binits{N.}},
\bauthor{\bsnm{Yang}, \binits{B.}},
\bauthor{\bsnm{Liu}, \binits{K.}},
\bauthor{\bsnm{Li}, \binits{H.}},
\bauthor{\bsnm{Chen}, \binits{G.}},
\bauthor{\bsnm{Qiu}, \binits{X.}},
\bauthor{\bsnm{Li}, \binits{W.}},
\bauthor{\bsnm{Hu}, \binits{J.}},
\bauthor{\bsnm{Fu}, \binits{J.}},
\bauthor{\bsnm{Jiang}, \binits{Y.}},
\bauthor{\bsnm{Liu}, \binits{M.}},
\bauthor{\bsnm{Ye}, \binits{J.}}:
\batitle{Machine learning in screening high performance electrocatalysts for {CO$_2$} reduction}.
\bjtitle{Small Methods}
\bvolume{5}(\bissue{11}),
\bfpage{2100987}
(\byear{2021})
\doiurl{10.1002/smtd.202100987}
\end{barticle}
\endbibitem

\bibitem[\protect\citeauthoryear{Lan et~al.}{2023}]{Lan2023}
\begin{barticle}
\bauthor{\bsnm{Lan}, \binits{J.}},
\bauthor{\bsnm{Palizhati}, \binits{A.}},
\bauthor{\bsnm{Shuaibi}, \binits{M.}}, \betal:
\batitle{{AdsorbML}: a leap in efficiency for adsorption energy calculations using generalizable machine learning potentials}.
\bjtitle{npj Comput Mater}
\bvolume{9},
\bfpage{172}
(\byear{2023})
\doiurl{10.1038/s41524-023-01121-5}
\end{barticle}
\endbibitem

\bibitem[\protect\citeauthoryear{Mou et~al.}{2023}]{mou2023machine}
\begin{barticle}
\bauthor{\bsnm{Mou}, \binits{L.-H.}},
\bauthor{\bsnm{Han}, \binits{T.}},
\bauthor{\bsnm{Smith}, \binits{P.E.S.}},
\bauthor{\bsnm{Sharman}, \binits{E.}},
\bauthor{\bsnm{Jiang}, \binits{J.}}:
\batitle{Machine learning descriptors for data-driven catalysis study}.
\bjtitle{Advanced Science}
\bvolume{10}(\bissue{22}),
\bfpage{2301020}
(\byear{2023})
\doiurl{10.1002/advs.202301020}
{\href{https://arxiv.org/abs/https://onlinelibrary.wiley.com/doi/pdf/10.1002/advs.202301020}{{https://onlinelibrary.wiley.com/doi/pdf/10.1002/advs.202301020}}}
\end{barticle}
\endbibitem

\bibitem[\protect\citeauthoryear{Zhang and Ling}{2018}]{Zhang2018strategy}
\begin{barticle}
\bauthor{\bsnm{Zhang}, \binits{Y.}},
\bauthor{\bsnm{Ling}, \binits{C.}}:
\batitle{A strategy to apply machine learning to small datasets in materials science}.
\bjtitle{npj Computational Materials}
\bvolume{4}(\bissue{1}),
\bfpage{25}
(\byear{2018})
\doiurl{10.1038/s41524-018-0081-z}
\end{barticle}
\endbibitem

\bibitem[\protect\citeauthoryear{Mamun et~al.}{2020}]{Mamun2020}
\begin{barticle}
\bauthor{\bsnm{Mamun}, \binits{O.}},
\bauthor{\bsnm{Winther}, \binits{K.T.}},
\bauthor{\bsnm{Boes}, \binits{J.R.}},
\bauthor{\bsnm{Bligaard}, \binits{T.}}:
\batitle{A bayesian framework for adsorption energy prediction on bimetallic alloy catalysts}.
\bjtitle{npj Computational Materials}
\bvolume{6}(\bissue{1}),
\bfpage{177}
(\byear{2020})
\doiurl{10.1038/s41524-020-00447-8}
\end{barticle}
\endbibitem

\bibitem[\protect\citeauthoryear{Fiedler et~al.}{2023}]{Fiedler2023}
\begin{barticle}
\bauthor{\bsnm{Fiedler}, \binits{L.}},
\bauthor{\bsnm{Modine}, \binits{N.A.}},
\bauthor{\bsnm{Schmerler}, \binits{S.}},
\bauthor{\bsnm{Vogel}, \binits{D.J.}},
\bauthor{\bsnm{Popoola}, \binits{G.A.}},
\bauthor{\bsnm{Thompson}, \binits{A.P.}},
\bauthor{\bsnm{Rajamanickam}, \binits{S.}},
\bauthor{\bsnm{Cangi}, \binits{A.}}:
\batitle{Predicting electronic structures at any length scale with machine learning}.
\bjtitle{npj Computational Materials}
\bvolume{9}(\bissue{1}),
\bfpage{115}
(\byear{2023})
\doiurl{10.1038/s41524-023-01070-z}
\end{barticle}
\endbibitem

\bibitem[\protect\citeauthoryear{Lunger et~al.}{2024}]{Lunger2024}
\begin{barticle}
\bauthor{\bsnm{Lunger}, \binits{J.R.}},
\bauthor{\bsnm{Karaguesian}, \binits{J.}},
\bauthor{\bsnm{Chun}, \binits{H.}}, \betal:
\batitle{Towards atom-level understanding of metal oxide catalysts for the oxygen evolution reaction with machine learning}.
\bjtitle{npj Computational Materials}
\bvolume{10},
\bfpage{80}
(\byear{2024})
\doiurl{10.1038/s41524-024-01273-y}
\end{barticle}
\endbibitem

\bibitem[\protect\citeauthoryear{Liu and Peng}{2024}]{LIU202425}
\begin{barticle}
\bauthor{\bsnm{Liu}, \binits{X.}},
\bauthor{\bsnm{Peng}, \binits{H.-J.}}:
\batitle{Toward next-generation heterogeneous catalysts: Empowering surface reactivity prediction with machine learning}.
\bjtitle{Engineering}
\bvolume{39},
\bfpage{25}--\blpage{44}
(\byear{2024})
\doiurl{10.1016/j.eng.2023.07.021}
\end{barticle}
\endbibitem

\bibitem[\protect\citeauthoryear{Kang et~al.}{2020}]{kang2020}
\begin{barticle}
\bauthor{\bsnm{Kang}, \binits{P.-L.}},
\bauthor{\bsnm{Shang}, \binits{C.}},
\bauthor{\bsnm{Liu}, \binits{Z.-P.}}:
\batitle{Large-scale atomic simulation via machine learning potentials constructed by global potential energy surface exploration}.
\bjtitle{Accounts of Chemical Research}
\bvolume{53}(\bissue{10}),
\bfpage{2119}--\blpage{2129}
(\byear{2020})
\doiurl{10.1021/acs.accounts.0c00472}
\end{barticle}
\endbibitem

\bibitem[\protect\citeauthoryear{Chen et~al.}{2023}]{chen2023}
\begin{barticle}
\bauthor{\bsnm{Chen}, \binits{D.}},
\bauthor{\bsnm{Shang}, \binits{C.}},
\bauthor{\bsnm{Liu}, \binits{Z.-P.}}:
\batitle{Machine-learning atomic simulation for heterogeneous catalysis}.
\bjtitle{npj Computational Materials}
\bvolume{9}(\bissue{1}),
\bfpage{2}
(\byear{2023})
\doiurl{10.1038/s41524-022-00959-5}
\end{barticle}
\endbibitem

\bibitem[\protect\citeauthoryear{Amann et~al.}{2022}]{amann2022}
\begin{barticle}
\bauthor{\bsnm{Amann}, \binits{P.}},
\bauthor{\bsnm{Klötzer}, \binits{B.}},
\bauthor{\bsnm{Degerman}, \binits{D.}},
\bauthor{\bsnm{Köpfle}, \binits{N.}},
\bauthor{\bsnm{Götsch}, \binits{T.}},
\bauthor{\bsnm{Lömker}, \binits{P.}},
\bauthor{\bsnm{Rameshan}, \binits{C.}},
\bauthor{\bsnm{Ploner}, \binits{K.}},
\bauthor{\bsnm{Bikaljevic}, \binits{D.}},
\bauthor{\bsnm{Wang}, \binits{H.-Y.}},
\bauthor{\bsnm{Soldemo}, \binits{M.}},
\bauthor{\bsnm{Shipilin}, \binits{M.}},
\bauthor{\bsnm{Goodwin}, \binits{C.M.}},
\bauthor{\bsnm{Gladh}, \binits{J.}},
\bauthor{\bsnm{Stenlid}, \binits{J.H.}},
\bauthor{\bsnm{Börner}, \binits{M.}},
\bauthor{\bsnm{Schlueter}, \binits{C.}},
\bauthor{\bsnm{Nilsson}, \binits{A.}}:
\batitle{The state of zinc in methanol synthesis over a {Zn/ZnO/Cu(211)} model catalyst}.
\bjtitle{Science}
\bvolume{376}(\bissue{6593}),
\bfpage{603}--\blpage{608}
(\byear{2022})
\doiurl{10.1126/science.abj7747}
{\href{https://arxiv.org/abs/https://www.science.org/doi/pdf/10.1126/science.abj7747}{{https://www.science.org/doi/pdf/10.1126/science.abj7747}}}
\end{barticle}
\endbibitem

\bibitem[\protect\citeauthoryear{Batchelor et~al.}{2019}]{batchelor2019high}
\begin{barticle}
\bauthor{\bsnm{Batchelor}, \binits{T.A.}},
\bauthor{\bsnm{Pedersen}, \binits{J.K.}},
\bauthor{\bsnm{Winther}, \binits{S.H.}},
\bauthor{\bsnm{Castelli}, \binits{I.E.}},
\bauthor{\bsnm{Jacobsen}, \binits{K.W.}},
\bauthor{\bsnm{Rossmeisl}, \binits{J.}}:
\batitle{High-entropy alloys as a discovery platform for electrocatalysis}.
\bjtitle{Joule}
\bvolume{3}(\bissue{3}),
\bfpage{834}--\blpage{845}
(\byear{2019})
\doiurl{10.1016/j.joule.2018.12.015}
\end{barticle}
\endbibitem

\bibitem[\protect\citeauthoryear{Pedersen et~al.}{2020}]{pedersen2020high}
\begin{barticle}
\bauthor{\bsnm{Pedersen}, \binits{J.K.}},
\bauthor{\bsnm{Batchelor}, \binits{T.A.}},
\bauthor{\bsnm{Bagger}, \binits{A.}},
\bauthor{\bsnm{Rossmeisl}, \binits{J.}}:
\batitle{High-entropy alloys as catalysts for the {CO$_2$} and {CO} reduction reactions}.
\bjtitle{ACS catalysis}
\bvolume{10}(\bissue{3}),
\bfpage{2169}--\blpage{2176}
(\byear{2020})
\doiurl{10.1021/acscatal.9b04343}
\end{barticle}
\endbibitem

\bibitem[\protect\citeauthoryear{Chanussot et~al.}{2021}]{chanussot2021open}
\begin{barticle}
\bauthor{\bsnm{Chanussot}, \binits{L.}},
\bauthor{\bsnm{Das}, \binits{A.}},
\bauthor{\bsnm{Goyal}, \binits{S.}},
\bauthor{\bsnm{Lavril}, \binits{T.}},
\bauthor{\bsnm{Shuaibi}, \binits{M.}},
\bauthor{\bsnm{Riviere}, \binits{M.}},
\bauthor{\bsnm{Tran}, \binits{K.}},
\bauthor{\bsnm{Heras-Domingo}, \binits{J.}},
\bauthor{\bsnm{Ho}, \binits{C.}},
\bauthor{\bsnm{Hu}, \binits{W.}},
\bauthor{\bsnm{Palizhati}, \binits{A.}},
\bauthor{\bsnm{Sriram}, \binits{A.}},
\bauthor{\bsnm{Wood}, \binits{B.}},
\bauthor{\bsnm{Yoon}, \binits{J.}},
\bauthor{\bsnm{Parikh}, \binits{D.}},
\bauthor{\bsnm{Zitnick}, \binits{C.L.}},
\bauthor{\bsnm{Ulissi}, \binits{Z.}}:
\batitle{Open {Ca}talyst 2020 {(OC20)} dataset and community challenges}.
\bjtitle{ACS Catalysis}
\bvolume{11}(\bissue{10}),
\bfpage{6059}--\blpage{6072}
(\byear{2021})
\doiurl{10.1021/acscatal.0c04525}
\end{barticle}
\endbibitem

\bibitem[\protect\citeauthoryear{Tran et~al.}{2023}]{tran2023open}
\begin{barticle}
\bauthor{\bsnm{Tran}, \binits{R.}},
\bauthor{\bsnm{Lan}, \binits{J.}},
\bauthor{\bsnm{Shuaibi}, \binits{M.}},
\bauthor{\bsnm{Wood}, \binits{B.M.}},
\bauthor{\bsnm{Goyal}, \binits{S.}},
\bauthor{\bsnm{Das}, \binits{A.}},
\bauthor{\bsnm{Heras-Domingo}, \binits{J.}},
\bauthor{\bsnm{Kolluru}, \binits{A.}},
\bauthor{\bsnm{Rizvi}, \binits{A.}},
\bauthor{\bsnm{Shoghi}, \binits{N.}},
\bauthor{\bsnm{Sriram}, \binits{A.}},
\bauthor{\bsnm{Therrien}, \binits{F.}},
\bauthor{\bsnm{Abed}, \binits{J.}},
\bauthor{\bsnm{Voznyy}, \binits{O.}},
\bauthor{\bsnm{Sargent}, \binits{E.H.}},
\bauthor{\bsnm{Ulissi}, \binits{Z.}},
\bauthor{\bsnm{Zitnick}, \binits{C.L.}}:
\batitle{{The Open Catalyst 2022 (OC22)} dataset and challenges for oxide electrocatalysts}.
\bjtitle{ACS Catal.}
\bvolume{13}(\bissue{5}),
\bfpage{3066}--\blpage{3084}
(\byear{2023})
\doiurl{10.1021/acscatal.2c05426}
\end{barticle}
\endbibitem

\bibitem[\protect\citeauthoryear{Ramdas et~al.}{2017}]{ramdas2017wasserstein}
\begin{barticle}
\bauthor{\bsnm{Ramdas}, \binits{A.}},
\bauthor{\bsnm{Garc{\'\i}a~Trillos}, \binits{N.}},
\bauthor{\bsnm{Cuturi}, \binits{M.}}:
\batitle{On {W}asserstein two-sample testing and related families of nonparametric tests}.
\bjtitle{Entropy}
\bvolume{19}(\bissue{2}),
\bfpage{47}
(\byear{2017})
\doiurl{10.3390/e19020047}
\end{barticle}
\endbibitem

\bibitem[\protect\citeauthoryear{Bahri et~al.}{2022}]{bahri2022meta-analysis}
\begin{barticle}
\bauthor{\bsnm{Bahri}, \binits{S.}},
\bauthor{\bsnm{Pathak}, \binits{S.}},
\bauthor{\bsnm{Singhahluwalia}, \binits{A.}},
\bauthor{\bsnm{Malav}, \binits{P.}},
\bauthor{\bsnm{Upadhyayula}, \binits{S.}}:
\batitle{Meta-analysis approach for understanding the characteristics of {CO$_2$} reduction catalysts for renewable fuel production}.
\bjtitle{Journal of Cleaner Production}
\bvolume{339},
\bfpage{130653}
(\byear{2022})
\doiurl{10.1016/j.jclepro.2022.130653}
\end{barticle}
\endbibitem

\bibitem[\protect\citeauthoryear{Jain et~al.}{2013}]{matproj2013}
\begin{barticle}
\bauthor{\bsnm{Jain}, \binits{A.}},
\bauthor{\bsnm{Ong}, \binits{S.P.}},
\bauthor{\bsnm{Hautier}, \binits{G.}},
\bauthor{\bsnm{Chen}, \binits{W.}},
\bauthor{\bsnm{Richards}, \binits{W.D.}},
\bauthor{\bsnm{Dacek}, \binits{S.}},
\bauthor{\bsnm{Cholia}, \binits{S.}},
\bauthor{\bsnm{Gunter}, \binits{D.}},
\bauthor{\bsnm{Skinner}, \binits{D.}},
\bauthor{\bsnm{Ceder}, \binits{G.}},
\bauthor{\bsnm{Persson}, \binits{K.A.}}:
\batitle{{Commentary: The Materials Project: A materials genome approach to accelerating materials innovation}}.
\bjtitle{APL Materials}
\bvolume{1}(\bissue{1}),
\bfpage{011002}
(\byear{2013})
\doiurl{10.1063/1.4812323}
{\href{https://arxiv.org/abs/https://pubs.aip.org/aip/apm/article-pdf/doi/10.1063/1.4812323/13163869/011002\_1\_online.pdf}{{https://pubs.aip.org/aip/apm/article-pdf/doi/10.1063/1.4812323/13163869/011002\_1\_online.pdf}}}
\end{barticle}
\endbibitem

\bibitem[\protect\citeauthoryear{Hammer et~al.}{1999}]{PhysRevB.59.7413}
\begin{barticle}
\bauthor{\bsnm{Hammer}, \binits{B.}},
\bauthor{\bsnm{Hansen}, \binits{L.B.}},
\bauthor{\bsnm{N\o{}rskov}, \binits{J.K.}}:
\batitle{Improved adsorption energetics within density-functional theory using revised {Perdew-Burke-Ernzerhof} functionals}.
\bjtitle{Phys. Rev. B}
\bvolume{59},
\bfpage{7413}--\blpage{7421}
(\byear{1999})
\doiurl{10.1103/PhysRevB.59.7413}
\end{barticle}
\endbibitem

\bibitem[\protect\citeauthoryear{Alam et~al.}{2021}]{D1CY00922B}
\begin{barticle}
\bauthor{\bsnm{Alam}, \binits{M.I.}},
\bauthor{\bsnm{Cheula}, \binits{R.}},
\bauthor{\bsnm{Moroni}, \binits{G.}},
\bauthor{\bsnm{Nardi}, \binits{L.}},
\bauthor{\bsnm{Maestri}, \binits{M.}}:
\batitle{Mechanistic and multiscale aspects of thermo-catalytic co$_2$ conversion to c1 products}.
\bjtitle{Catal. Sci. Technol.}
\bvolume{11},
\bfpage{6601}--\blpage{6629}
(\byear{2021})
\doiurl{10.1039/D1CY00922B}
\end{barticle}
\endbibitem

\bibitem[\protect\citeauthoryear{Wu et~al.}{2017}]{Wu2017}
\begin{barticle}
\bauthor{\bsnm{Wu}, \binits{Z.}},
\bauthor{\bsnm{Cole}, \binits{J.}},
\bauthor{\bsnm{Fang}, \binits{H.L.}},
\bauthor{\bsnm{Qin}, \binits{M.}},
\bauthor{\bsnm{He}, \binits{Z.}}:
\batitle{Revisiting catalyst structure and mechanism in methanol synthesis}.
\bjtitle{Journal of Advances in Nanomaterials}
\bvolume{2}(\bissue{1}),
\bfpage{1}--\blpage{10}
(\byear{2017})
\doiurl{10.22606/jan.2017.21001}
\end{barticle}
\endbibitem

\bibitem[\protect\citeauthoryear{FAIR-Chem}{2024}]{fairchem}
\begin{botherref}
\oauthor{\bsnm{FAIR-Chem}}:
fairchem: A FAIR-Chem Project Repository.
GitHub
(2024).
\url{https://github.com/FAIR-Chem/fairchem}
\end{botherref}
\endbibitem

\bibitem[\protect\citeauthoryear{Liao et~al.}{2023}]{liao2023equiformerv2}
\begin{barticle}
\bauthor{\bsnm{Liao}, \binits{Y.-L.}},
\bauthor{\bsnm{Wood}, \binits{B.}},
\bauthor{\bsnm{Das}, \binits{A.}},
\bauthor{\bsnm{Smidt}, \binits{T.}}:
\batitle{Equiformerv2: Improved equivariant transformer for scaling to higher-degree representations}.
\bjtitle{arXiv preprint arXiv:2306.12059}
(\byear{2023})
\doiurl{10.48550/arXiv.2306.12059} .
\bcomment{Published as a conference paper at ICLR 2024}
\end{barticle}
\endbibitem

\bibitem[\protect\citeauthoryear{Pisal and Krejci}{2024}]{dataset}
\begin{botherref}
\oauthor{\bsnm{Pisal}, \binits{P.}},
\oauthor{\bsnm{Krejci}, \binits{O.}}:
{Final Geometries and Energies, Statistical Analysis and Estimated Errors of Single Metals and Bimetallics for {CO$_2$} to Methanol Conversion}.
\doiurl{10.5281/zenodo.13370374} .
\url{https://doi.org/10.5281/zenodo.13370374}
\end{botherref}
\endbibitem

\bibitem[\protect\citeauthoryear{Jr.}{1963}]{ward1963hierarchical}
\begin{barticle}
\bauthor{\bsnm{Jr.}, \binits{J.H.W.}}:
\batitle{Hierarchical grouping to optimize an objective function}.
\bjtitle{Journal of the American Statistical Association}
\bvolume{58}(\bissue{301}),
\bfpage{236}--\blpage{244}
(\byear{1963})
\doiurl{10.1080/01621459.1963.10500845}
\end{barticle}
\endbibitem

\bibitem[\protect\citeauthoryear{Dongapure et~al.}{2023}]{Dongapure_ChemCatChem23_NiZn}
\begin{barticle}
\bauthor{\bsnm{Dongapure}, \binits{P.}},
\bauthor{\bsnm{Tekawadia}, \binits{J.}},
\bauthor{\bsnm{Thundiyil}, \binits{S.}},
\bauthor{\bsnm{Caha}, \binits{I.}},
\bauthor{\bsnm{Deepak}, \binits{F.L.}},
\bauthor{\bsnm{Devi}, \binits{R.N.}}:
\batitle{Mechanistic insights into near ambient pressure activity of intermetallic {NiZn/TiO$_2$} catalyst for {CO$_2$} conversion to methanol}.
\bjtitle{ChemCatChem}
\bvolume{15}(\bissue{1}),
\bfpage{202201150}
(\byear{2023})
\doiurl{10.1002/cctc.202201150}
{\href{https://arxiv.org/abs/https://chemistry-europe.onlinelibrary.wiley.com/doi/pdf/10.1002/cctc.202201150}{{https://chemistry-europe.onlinelibrary.wiley.com/doi/pdf/10.1002/cctc.202201150}}}
\end{barticle}
\endbibitem

\bibitem[\protect\citeauthoryear{Beck et~al.}{2024}]{Beck2024}
\begin{barticle}
\bauthor{\bsnm{Beck}, \binits{A.}},
\bauthor{\bsnm{Newton}, \binits{M.A.}},
\bauthor{\bsnm{Water}, \binits{L.G.A.}},
\bauthor{\bsnm{Bokhoven}, \binits{J.A.}}:
\batitle{The enigma of methanol synthesis by cu/zno/al2o3-based catalysts}.
\bjtitle{Chemical Reviews}
\bvolume{124}(\bissue{8}),
\bfpage{4543}--\blpage{4678}
(\byear{2024})
\doiurl{10.1021/acs.chemrev.3c00148}
\end{barticle}
\endbibitem

\bibitem[\protect\citeauthoryear{Behrens et~al.}{2012}]{BehrensCu211Sci21}
\begin{barticle}
\bauthor{\bsnm{Behrens}, \binits{M.}},
\bauthor{\bsnm{Studt}, \binits{F.}},
\bauthor{\bsnm{Kasatkin}, \binits{I.}},
\bauthor{\bsnm{Kühl}, \binits{S.}},
\bauthor{\bsnm{Hävecker}, \binits{M.}},
\bauthor{\bsnm{Abild-Pedersen}, \binits{F.}},
\bauthor{\bsnm{Zander}, \binits{S.}},
\bauthor{\bsnm{Girgsdies}, \binits{F.}},
\bauthor{\bsnm{Kurr}, \binits{P.}},
\bauthor{\bsnm{Kniep}, \binits{B.-L.}},
\bauthor{\bsnm{Tovar}, \binits{M.}},
\bauthor{\bsnm{Fischer}, \binits{R.W.}},
\bauthor{\bsnm{Nørskov}, \binits{J.K.}},
\bauthor{\bsnm{Schlögl}, \binits{R.}}:
\batitle{The active site of methanol synthesis over {Cu/ZnO/Al$_2$O$_3$} industrial catalysts}.
\bjtitle{Science}
\bvolume{336}(\bissue{6083}),
\bfpage{893}--\blpage{897}
(\byear{2012})
\doiurl{10.1126/science.1219831}
{\href{https://arxiv.org/abs/https://www.science.org/doi/pdf/10.1126/science.1219831}{{https://www.science.org/doi/pdf/10.1126/science.1219831}}}
\end{barticle}
\endbibitem

\bibitem[\protect\citeauthoryear{Laudenschleger et~al.}{2020}]{Laudenschleger2020}
\begin{barticle}
\bauthor{\bsnm{Laudenschleger}, \binits{D.}},
\bauthor{\bsnm{Ruland}, \binits{H.}},
\bauthor{\bsnm{Muhler}, \binits{M.}}:
\batitle{Identifying the nature of the active sites in methanol synthesis over cu/zno/al2o3 catalysts}.
\bjtitle{Nature Communications}
\bvolume{11}(\bissue{1}),
\bfpage{3898}
(\byear{2020})
\doiurl{10.1038/s41467-020-17631-5}
\end{barticle}
\endbibitem

\bibitem[\protect\citeauthoryear{Studt et~al.}{2014}]{Studt2014GaNi}
\begin{barticle}
\bauthor{\bsnm{Studt}, \binits{F.}},
\bauthor{\bsnm{Sharafutdinov}, \binits{I.}},
\bauthor{\bsnm{Abild-Pedersen}, \binits{F.}},
\bauthor{\bsnm{Elkjær}, \binits{C.F.}},
\bauthor{\bsnm{Hummelshøj}, \binits{J.S.}},
\bauthor{\bsnm{Dahl}, \binits{S.}},
\bauthor{\bsnm{Chorkendorff}, \binits{I.}},
\bauthor{\bsnm{Nørskov}, \binits{J.K.}}:
\batitle{Discovery of a {Ni-Ga} catalyst for carbon dioxide reduction to methanol}.
\bjtitle{Nature Chemistry}
\bvolume{6}(\bissue{4}),
\bfpage{320}
(\byear{2014})
\doiurl{10.1038/NCHEM.1873}
\end{barticle}
\endbibitem

\bibitem[\protect\citeauthoryear{Toyir et~al.}{2001}]{CuGaTOYIR01}
\begin{barticle}
\bauthor{\bsnm{Toyir}, \binits{J.}},
\bauthor{\bsnm{{Ramirez de la Piscina}}, \binits{P.}},
\bauthor{\bsnm{Fierro}, \binits{J.L.G.}},
\bauthor{\bsnm{Homs}, \binits{N.}}:
\batitle{Catalytic performance for co$_2$ conversion to methanol of gallium-promoted copper-based catalysts: influence of metallic precursors}.
\bjtitle{Applied Catalysis B: Environmental}
\bvolume{34}(\bissue{4}),
\bfpage{255}--\blpage{266}
(\byear{2001})
\doiurl{10.1016/S0926-3373(01)00203-X}
\end{barticle}
\endbibitem

\bibitem[\protect\citeauthoryear{Medina et~al.}{2017}]{Medina2017_GaCu}
\begin{barticle}
\bauthor{\bsnm{Medina}, \binits{J.C.}},
\bauthor{\bsnm{Figueroa}, \binits{M.}},
\bauthor{\bsnm{Manrique}, \binits{R.}},
\bauthor{\bsnm{Rodríguez~Pereira}, \binits{J.}},
\bauthor{\bsnm{Srinivasan}, \binits{P.D.}},
\bauthor{\bsnm{Bravo-Suárez}, \binits{J.J.}},
\bauthor{\bsnm{Baldovino~Medrano}, \binits{V.G.}},
\bauthor{\bsnm{Jiménez}, \binits{R.}},
\bauthor{\bsnm{Karelovic}, \binits{A.}}:
\batitle{Catalytic consequences of ga promotion on cu for co$_2$ hydrogenation to methanol}.
\bjtitle{Catalysis Science \& Technology}
\bvolume{7}(\bissue{15}),
\bfpage{3375}--\blpage{3387}
(\byear{2017})
\doiurl{10.1039/C7CY01021D}
\end{barticle}
\endbibitem

\bibitem[\protect\citeauthoryear{Men et~al.}{2019}]{PtInMen2019}
\begin{barticle}
\bauthor{\bsnm{Men}, \binits{Y.-L.}},
\bauthor{\bsnm{Liu}, \binits{Y.}},
\bauthor{\bsnm{Wang}, \binits{Q.}},
\bauthor{\bsnm{Luo}, \binits{Z.-H.}},
\bauthor{\bsnm{Shao}, \binits{S.}},
\bauthor{\bsnm{Li}, \binits{Y.-B.}},
\bauthor{\bsnm{Pan}, \binits{Y.-X.}}:
\batitle{Highly dispersed pt-based catalysts for selective co$_2$ hydrogenation to methanol at atmospheric pressure}.
\bjtitle{Chemical Engineering Science}
\bvolume{200},
\bfpage{167}--\blpage{175}
(\byear{2019})
\doiurl{10.1016/j.ces.2019.02.004}
\end{barticle}
\endbibitem

\bibitem[\protect\citeauthoryear{Tang et~al.}{2024}]{Tang2024}
\begin{barticle}
\bauthor{\bsnm{Tang}, \binits{X.}},
\bauthor{\bsnm{Song}, \binits{C.}},
\bauthor{\bsnm{Li}, \binits{H.}}, \betal:
\batitle{Thermally stable ni foam-supported inverse {CeAlO$_x$/Ni} ensemble as an active structured catalyst for {CO$_2$} hydrogenation to methane}.
\bjtitle{Nat Commun}
\bvolume{15}(\bissue{3115}),
\bfpage{3115}
(\byear{2024})
\doiurl{10.1038/s41467-024-47403-4}
\end{barticle}
\endbibitem

\bibitem[\protect\citeauthoryear{Hu et~al.}{2022}]{HuNiBasedACS22}
\begin{barticle}
\bauthor{\bsnm{Hu}, \binits{F.}},
\bauthor{\bsnm{Ye}, \binits{R.}},
\bauthor{\bsnm{Lu}, \binits{Z.-H.}},
\bauthor{\bsnm{Zhang}, \binits{R.}},
\bauthor{\bsnm{Feng}, \binits{G.}}:
\batitle{Structure–activity relationship of {Ni}-based catalysts toward {CO$_2$} methanation: Recent advances and future perspectives}.
\bjtitle{Energy \& Fuels}
\bvolume{36}(\bissue{1}),
\bfpage{156}--\blpage{169}
(\byear{2022})
\doiurl{10.1021/acs.energyfuels.1c03645}
{\href{https://arxiv.org/abs/https://doi.org/10.1021/acs.energyfuels.1c03645}{{https://doi.org/10.1021/acs.energyfuels.1c03645}}}
\end{barticle}
\endbibitem

\bibitem[\protect\citeauthoryear{Lempelto et~al.}{2023}]{lempelto2023}
\begin{barticle}
\bauthor{\bsnm{Lempelto}, \binits{A.}},
\bauthor{\bsnm{Gell}, \binits{L.}},
\bauthor{\bsnm{Kiljunen}, \binits{T.}},
\bauthor{\bsnm{Honkala}, \binits{K.}}:
\batitle{Exploring co$_2$ hydrogenation to methanol at a cuzn–zro2 interface via dft calculations}.
\bjtitle{Catal. Sci. Technol.}
\bvolume{13},
\bfpage{4387}--\blpage{4399}
(\byear{2023})
\doiurl{10.1039/D3CY00549F}
\end{barticle}
\endbibitem

\bibitem[\protect\citeauthoryear{Wang et~al.}{2019}]{Wang2019_MaZrOx}
\begin{barticle}
\bauthor{\bsnm{Wang}, \binits{J.}},
\bauthor{\bsnm{Tang}, \binits{C.}},
\bauthor{\bsnm{Li}, \binits{G.}},
\bauthor{\bsnm{Han}, \binits{Z.}},
\bauthor{\bsnm{Li}, \binits{Z.}},
\bauthor{\bsnm{Liu}, \binits{H.}},
\bauthor{\bsnm{Cheng}, \binits{F.}},
\bauthor{\bsnm{Li}, \binits{C.}}:
\batitle{High-performance mazrox (ma = cd, ga) solid-solution catalysts for co$_2$ hydrogenation to methanol}.
\bjtitle{ACS Catalysis}
\bvolume{9}(\bissue{11}),
\bfpage{10253}--\blpage{10259}
(\byear{2019})
\doiurl{10.1021/acscatal.9b03449}
\end{barticle}
\endbibitem

\bibitem[\protect\citeauthoryear{Cheula et~al.}{2024}]{Cheula2024_DopantsZirconia}
\begin{barticle}
\bauthor{\bsnm{Cheula}, \binits{R.}},
\bauthor{\bsnm{Tran}, \binits{T.A.M.Q.}},
\bauthor{\bsnm{Andersen}, \binits{M.}}:
\batitle{Unraveling the effect of dopants in zirconia-based catalysts for co$_2$ hydrogenation to methanol}.
\bjtitle{ACS Catalysis}
\bvolume{14}(\bissue{17}),
\bfpage{13126}--\blpage{13135}
(\byear{2024})
\doiurl{10.1021/acscatal.4c03206}
\end{barticle}
\endbibitem

\bibitem[\protect\citeauthoryear{Cannizzaro et~al.}{2023}]{Cannizzaro2023}
\begin{barticle}
\bauthor{\bsnm{Cannizzaro}, \binits{F.}},
\bauthor{\bsnm{Hensen}, \binits{E.J.M.}},
\bauthor{\bsnm{Filot}, \binits{I.A.W.}}:
\batitle{The promoting role of ni on in2o3 for co$_2$ hydrogenation to methanol}.
\bjtitle{ACS Catalysis}
\bvolume{13}(\bissue{3}),
\bfpage{1875}--\blpage{1892}
(\byear{2023})
\doiurl{10.1021/acscatal.2c04872}
\end{barticle}
\endbibitem

\bibitem[\protect\citeauthoryear{Gao et~al.}{2022}]{Gao2022}
\begin{barticle}
\bauthor{\bsnm{Gao}, \binits{D.}},
\bauthor{\bsnm{Li}, \binits{W.}},
\bauthor{\bsnm{Wang}, \binits{H.}},
\bauthor{\bsnm{Wang}, \binits{G.}},
\bauthor{\bsnm{Cai}, \binits{R.}}:
\batitle{Heterogeneous catalysis for {CO$_2$} conversion into chemicals and fuels}.
\bjtitle{Transactions of Tianjin University}
\bvolume{28}(\bissue{4}),
\bfpage{245}--\blpage{264}
(\byear{2022})
\doiurl{10.1007/s12209-022-00326-x}
\end{barticle}
\endbibitem

\bibitem[\protect\citeauthoryear{Liu et~al.}{2024}]{LIU2024NiZr}
\begin{barticle}
\bauthor{\bsnm{Liu}, \binits{L.}},
\bauthor{\bsnm{Gao}, \binits{Y.}},
\bauthor{\bsnm{Zhang}, \binits{H.}},
\bauthor{\bsnm{Kosinov}, \binits{N.}},
\bauthor{\bsnm{Hensen}, \binits{E.J.M.}}:
\batitle{Ni and zro2 promotion of in2o3 for co$_2$ hydrogenation to methanol}.
\bjtitle{Applied Catalysis B: Environment and Energy}
\bvolume{356},
\bfpage{124210}
(\byear{2024})
\doiurl{10.1016/j.apcatb.2024.124210}
\end{barticle}
\endbibitem

\bibitem[\protect\citeauthoryear{Kresse and Hafner}{1994}]{vasp1_PhysRevB.49.14251}
\begin{barticle}
\bauthor{\bsnm{Kresse}, \binits{G.}},
\bauthor{\bsnm{Hafner}, \binits{J.}}:
\batitle{Ab initio molecular-dynamics simulation of the liquid-metal--amorphous-semiconductor transition in germanium}.
\bjtitle{Phys. Rev. B}
\bvolume{49},
\bfpage{14251}--\blpage{14269}
(\byear{1994})
\doiurl{10.1103/PhysRevB.49.14251}
\end{barticle}
\endbibitem

\bibitem[\protect\citeauthoryear{Kresse and Furthm\"uller}{1996}]{vasp2_PhysRevB.54.11169}
\begin{barticle}
\bauthor{\bsnm{Kresse}, \binits{G.}},
\bauthor{\bsnm{Furthm\"uller}, \binits{J.}}:
\batitle{Efficient iterative schemes for ab initio total-energy calculations using a plane-wave basis set}.
\bjtitle{Phys. Rev. B}
\bvolume{54},
\bfpage{11169}--\blpage{11186}
(\byear{1996})
\doiurl{10.1103/PhysRevB.54.11169}
\end{barticle}
\endbibitem

\bibitem[\protect\citeauthoryear{Mathew et~al.}{2017}]{atomate_2017}
\begin{barticle}
\bauthor{\bsnm{Mathew}, \binits{K.}},
\bauthor{\bsnm{Montoya}, \binits{J.H.}},
\bauthor{\bsnm{Faghaninia}, \binits{A.}},
\bauthor{\bsnm{Dwarakanath}, \binits{S.}},
\bauthor{\bsnm{Aykol}, \binits{M.}},
\bauthor{\bsnm{Tang}, \binits{H.}},
\bauthor{\bsnm{Chu}, \binits{I.-h.}},
\bauthor{\bsnm{Smidt}, \binits{T.}},
\bauthor{\bsnm{Bocklund}, \binits{B.}},
\bauthor{\bsnm{Horton}, \binits{M.}},
\bauthor{\bsnm{Dagdelen}, \binits{J.}},
\bauthor{\bsnm{Wood}, \binits{B.}},
\bauthor{\bsnm{Liu}, \binits{Z.-K.}},
\bauthor{\bsnm{Neaton}, \binits{J.}},
\bauthor{\bsnm{Ong}, \binits{S.P.}},
\bauthor{\bsnm{Persson}, \binits{K.}},
\bauthor{\bsnm{Jain}, \binits{A.}}:
\batitle{Atomate: A high-level interface to generate, execute, and analyze computational materials science workflows}.
\bjtitle{Computational Materials Science}
\bvolume{139},
\bfpage{140}--\blpage{152}
(\byear{2017})
\doiurl{10.1016/j.commatsci.2017.07.030}
\end{barticle}
\endbibitem

\bibitem[\protect\citeauthoryear{Ong et~al.}{2013a}]{pymatgen_2013}
\begin{barticle}
\bauthor{\bsnm{Ong}, \binits{S.P.}},
\bauthor{\bsnm{Richards}, \binits{W.D.}},
\bauthor{\bsnm{Jain}, \binits{A.}},
\bauthor{\bsnm{Hautier}, \binits{G.}},
\bauthor{\bsnm{Kocher}, \binits{M.}},
\bauthor{\bsnm{Cholia}, \binits{S.}},
\bauthor{\bsnm{Gunter}, \binits{D.}},
\bauthor{\bsnm{Chevrier}, \binits{V.L.}},
\bauthor{\bsnm{Persson}, \binits{K.A.}},
\bauthor{\bsnm{Ceder}, \binits{G.}}:
\batitle{Python materials genomics (pymatgen): A robust, open-source python library for materials analysis}.
\bjtitle{Computational Materials Science}
\bvolume{68},
\bfpage{314}--\blpage{319}
(\byear{2013})
\doiurl{10.1016/j.commatsci.2012.10.028}
\end{barticle}
\endbibitem

\bibitem[\protect\citeauthoryear{Ong et~al.}{2013b}]{custodian_2013}
\begin{barticle}
\bauthor{\bsnm{Ong}, \binits{S.P.}},
\bauthor{\bsnm{Richards}, \binits{W.D.}},
\bauthor{\bsnm{Jain}, \binits{A.}},
\bauthor{\bsnm{Hautier}, \binits{G.}},
\bauthor{\bsnm{Kocher}, \binits{M.}},
\bauthor{\bsnm{Cholia}, \binits{S.}},
\bauthor{\bsnm{Gunter}, \binits{D.}},
\bauthor{\bsnm{Chevrier}, \binits{V.L.}},
\bauthor{\bsnm{Persson}, \binits{K.A.}},
\bauthor{\bsnm{Ceder}, \binits{G.}}:
\batitle{Python materials genomics (pymatgen): A robust, open-source python library for materials analysis}.
\bjtitle{Computational Materials Science}
\bvolume{68},
\bfpage{314}--\blpage{319}
(\byear{2013})
\doiurl{10.1016/j.commatsci.2012.10.028}
\end{barticle}
\endbibitem

\bibitem[\protect\citeauthoryear{Jain et~al.}{2015}]{firworks_2015}
\begin{barticle}
\bauthor{\bsnm{Jain}, \binits{A.}},
\bauthor{\bsnm{Ong}, \binits{S.P.}},
\bauthor{\bsnm{Chen}, \binits{W.}},
\bauthor{\bsnm{Medasani}, \binits{B.}},
\bauthor{\bsnm{Qu}, \binits{X.}},
\bauthor{\bsnm{Kocher}, \binits{M.}},
\bauthor{\bsnm{Brafman}, \binits{M.}},
\bauthor{\bsnm{Petretto}, \binits{G.}},
\bauthor{\bsnm{Rignanese}, \binits{G.-M.}},
\bauthor{\bsnm{Hautier}, \binits{G.}},
\bauthor{\bsnm{Gunter}, \binits{D.}},
\bauthor{\bsnm{Persson}, \binits{K.A.}}:
\batitle{Fireworks: a dynamic workflow system designed for high-throughput applications}.
\bjtitle{Concurrency and Computation: Practice and Experience}
\bvolume{27}(\bissue{17}),
\bfpage{5037}--\blpage{5059}
(\byear{2015})
\doiurl{10.1002/cpe.3505}
{\href{https://arxiv.org/abs/https://onlinelibrary.wiley.com/doi/pdf/10.1002/cpe.3505}{{https://onlinelibrary.wiley.com/doi/pdf/10.1002/cpe.3505}}}
\end{barticle}
\endbibitem

\bibitem[\protect\citeauthoryear{Gasteiger et~al.}{2022}]{gasteiger2022gemnetocdevelopinggraphneural}
\begin{botherref}
\oauthor{\bsnm{Gasteiger}, \binits{J.}},
\oauthor{\bsnm{Shuaibi}, \binits{M.}},
\oauthor{\bsnm{Sriram}, \binits{A.}},
\oauthor{\bsnm{Günnemann}, \binits{S.}},
\oauthor{\bsnm{Ulissi}, \binits{Z.}},
\oauthor{\bsnm{Zitnick}, \binits{C.L.}},
\oauthor{\bsnm{Das}, \binits{A.}}:
GemNet-OC: Developing Graph Neural Networks for Large and Diverse Molecular Simulation Datasets
(2022).
\url{https://arxiv.org/abs/2204.02782}
\end{botherref}
\endbibitem

\bibitem[\protect\citeauthoryear{Virtanen et~al.}{2020}]{2020SciPy-NMeth}
\begin{barticle}
\bauthor{\bsnm{Virtanen}, \binits{P.}},
\bauthor{\bsnm{Gommers}, \binits{R.}},
\bauthor{\bsnm{Oliphant}, \binits{T.E.}},
\bauthor{\bsnm{Haberland}, \binits{M.}},
\bauthor{\bsnm{Reddy}, \binits{T.}},
\bauthor{\bsnm{Cournapeau}, \binits{D.}},
\bauthor{\bsnm{Burovski}, \binits{E.}},
\bauthor{\bsnm{Peterson}, \binits{P.}},
\bauthor{\bsnm{Weckesser}, \binits{W.}},
\bauthor{\bsnm{Bright}, \binits{J.}},
\bauthor{\bsnm{{van der Walt}}, \binits{S.J.}},
\bauthor{\bsnm{Brett}, \binits{M.}},
\bauthor{\bsnm{Wilson}, \binits{J.}},
\bauthor{\bsnm{Millman}, \binits{K.J.}},
\bauthor{\bsnm{Mayorov}, \binits{N.}},
\bauthor{\bsnm{Nelson}, \binits{A.R.J.}},
\bauthor{\bsnm{Jones}, \binits{E.}},
\bauthor{\bsnm{Kern}, \binits{R.}},
\bauthor{\bsnm{Larson}, \binits{E.}},
\bauthor{\bsnm{Carey}, \binits{C.J.}},
\bauthor{\bsnm{Polat}, \binits{{\. I}.}},
\bauthor{\bsnm{Feng}, \binits{Y.}},
\bauthor{\bsnm{Moore}, \binits{E.W.}},
\bauthor{\bsnm{{VanderPlas}}, \binits{J.}},
\bauthor{\bsnm{Laxalde}, \binits{D.}},
\bauthor{\bsnm{Perktold}, \binits{J.}},
\bauthor{\bsnm{Cimrman}, \binits{R.}},
\bauthor{\bsnm{Henriksen}, \binits{I.}},
\bauthor{\bsnm{Quintero}, \binits{E.A.}},
\bauthor{\bsnm{Harris}, \binits{C.R.}},
\bauthor{\bsnm{Archibald}, \binits{A.M.}},
\bauthor{\bsnm{Ribeiro}, \binits{A.H.}},
\bauthor{\bsnm{Pedregosa}, \binits{F.}},
\bauthor{\bsnm{{van Mulbregt}}, \binits{P.}},
\bauthor{\bsnm{{SciPy 1.0 Contributors}}}:
\batitle{{{SciPy} 1.0: Fundamental Algorithms for Scientific Computing in Python}}.
\bjtitle{Nature Methods}
\bvolume{17},
\bfpage{261}--\blpage{272}
(\byear{2020})
\doiurl{10.1038/s41592-019-0686-2}
\end{barticle}
\endbibitem

\bibitem[\protect\citeauthoryear{Müllner}{2011}]{müllner2011modernhierarchicalagglomerativeclustering}
\begin{botherref}
\oauthor{\bsnm{Müllner}, \binits{D.}}:
Modern hierarchical, agglomerative clustering algorithms
(2011).
\url{https://arxiv.org/abs/1109.2378}
\end{botherref}
\endbibitem

\bibitem[\protect\citeauthoryear{Bar-Joseph et~al.}{2001}]{bar2001hierarchical}
\begin{barticle}
\bauthor{\bsnm{Bar-Joseph}, \binits{Z.}},
\bauthor{\bsnm{Gifford}, \binits{D.K.}},
\bauthor{\bsnm{Jaakkola}, \binits{T.S.}}:
\batitle{Fast optimal leaf ordering for hierarchical clustering}.
\bjtitle{Bioinformatics}
\bvolume{17}(\bissue{1}),
\bfpage{22}--\blpage{29}
(\byear{2001})
\doiurl{10.1093/bioinformatics/17.suppl_1.S22}
{\href{https://arxiv.org/abs/https://academic.oup.com/bioinformatics/article-pdf/17/suppl\_1/S22/50522365/bioinformatics\_17\_suppl1\_s22.pdf}{{https://academic.oup.com/bioinformatics/article-pdf/17/suppl\_1/S22/50522365/bioinformatics\_17\_suppl1\_s22.pdf}}}
\end{barticle}
\endbibitem

\end{thebibliography}

\newpage

\newpage

\end{document}